\documentclass[acmlarge]{acmart}
\makeatletter
\newcommand{\myconfshort}{\acmConference@shortname}
\newcommand{\myconffull}{\acmConference@name}
\newcommand{\myconfdate}{\acmConference@date}
\newcommand{\myconfloc}{\acmConference@venue}
\AtBeginDocument{
  \fancypagestyle{firstpagestyle}{
    \fancyhead{}%
    \fancyfoot[C]{}%
  }
  \fancyhf{}
  \fancyhead[LO]{\@headfootfont\shorttitle}%
  \fancyhead[RE]{\@headfootfont\@shortauthors}%
  \fancyhead[LE]{\@headfootfont\footnotesize \myconfshort, \myconfdate, \myconfloc}%
  \fancyhead[RO]{\@headfootfont\footnotesize \myconfshort, \myconfdate, \myconfloc}%
  \fancyfoot[C]{}%
}
\makeatother
\acmBooktitle{\conffull\@ (\confshort), \confdate, \confloc}
\AtBeginDocument{%
  }

\setcopyright{acmlicensed}
\copyrightyear{2026}
\acmYear{2026}
\setcopyright{cc}
\setcctype{by}
\acmConference[FAccT '26]{The 2026 ACM Conference on Fairness, Accountability, and Transparency}{June 25--28, 2026}{Montreal, QC, Canada}
\acmBooktitle{The 2026 ACM Conference on Fairness, Accountability, and Transparency (FAccT '26), June 25--28, 2026, Montreal, QC, Canada}
\acmDOI{10.1145/3805689.3812334}
\acmISBN{979-8-4007-2596-8/2026/06}
\acmISBN{978-1-4503-XXXX-X/2018/06}

\usepackage{multirow}
\usepackage[normalem]{ulem}

\newcommand{\modelname}[1]{\textsf{#1}} 

\begin{document}

\title{Informing AI Policy Assessment using Large-Scale Simulation of Interventions}

\author{Julia Barnett}
\email{JuliaBarnett@u.northwestern.edu}
\orcid{0000-0002-3476-1110}
\affiliation{%
  \institution{Northwestern University}
  \country{USA}
}
\author{Kimon Kieslich}
\affiliation{
  \institution{University of Amsterdam, The Netherlands \& University of Hohenheim}
  \country{Germany}
}
\orcid{0000-0002-6305-2997}

\author{Natali Helberger}
\affiliation{%
  \institution{University of Amsterdam}
  \country{The Netherlands}
}
\orcid{0000-0003-1652-0580}

\author{Nicholas Diakopoulos}
\affiliation{%
 \institution{Northwestern University}
 \country{USA}}
\orcid{0000-0001-5005-6123}

\renewcommand{\shortauthors}{Barnett et al.}
\renewcommand{\shorttitle}{Informing AI Policy Assessment}

\begin{abstract}
  As the rapid proliferation of AI systems and harms spurs efforts in AI governance around the world, prioritizing among competing policy options has become increasingly challenging for policymakers and researchers. We introduce a methodology for identifying viable policy options to mitigate specified AI harms, helping policymakers and researchers target areas that warrant greater time and resource investment. This method combines participatory evaluation of policies, expert assessment of implementation costs, and an LLM-based assessment of perceived harm mitigation under each policy option. We leverage a genetic algorithm-based simulation study to explore a vast solution space of potential policy combinations, and examine how outcomes change under different weightings of cost, participatory input, and harm mitigation. We find that this method enables exploration of different balances between participatory and expert components, allowing policymakers and researchers to assess how much weight to assign to each. We argue that the diversity of viable policy combinations found by the genetic algorithm could be a useful starting point for deliberation. This method operationalizes existing work on participatory AI by integrating it directly into practical policy development pipelines.
\end{abstract}


\received{13 January 2026}

\maketitle

\section{Introduction}



Given the increasing impact AI systems have on individuals and society, governing harms has become a challenge for policymakers and researchers alike \cite{davtyan2024overview}. The speed of AI implementation and the multitude of purposes and contexts for AI use has resulted in a rapid expansion of AI governance 
around the world \cite{roberts2024global} from the national and supra-national to the local. Whether it be calls for more transparency about data and models, risk assessment protocols for developers, or labeling of outputs, there is a growing number of potential policy options that call for specific stakeholders to take specific actions to mitigate or prevent adverse outcomes for individuals and society. 

This multitude of options makes it difficult to identify and decide which strategies to prioritize when attempting to mitigate the risks of AI. In developing policies it is helpful to understand the different impacts of various policy options prior to getting to more expensive evaluations like randomized control trials \cite{petticrew2012rct} in order to orient policy decision makers. Policy assessment can be used to understand impacts of policy early on in the decision making process \cite{Adelle01032012, weimer2017policy}. 
As regulatory bodies and technologists continue to develop different AI governance methods, inclusive policy assessment is increasingly needed to evaluate impacts different individuals, organizations, and society at large. Many existing AI governance approaches rely on expert knowledge and do not involve a broader participative base, resulting in biased risk assessment \cite{kieslich2026scenario, orwat2024normative, reuel2025evaluates}. In this work we incorporate a perspective from \textit{lay stakeholders}, those who experience potential harms from AI systems in their daily lives and thus possess a contextual and situated expertise, but lack the technical expertise or institutional power to influence the systems causing them. Though some approaches have integrated more diverse stakeholders in this decision making \cite{mun2024particip, kieslich2025anticipating, barnett2025envisioning}, little has been done to show how these can complement expert knowledge, 
especially considering competing factors such as anticipated effectiveness, cost, 
and political viability of these actions. 
Policymakers are faced with an abundance of options to mitigate AI harms, all weighed differently across these factors, making it difficult to 
prioritize a balanced set of policy measures. 

In response, we \textbf{propose a method} that supports policy assessment by \textbf{identifying viable policy options} designed to mitigate harms introduced by AI and help policymakers identify areas where allocating time and resources for further assessment may be warranted. Our method provides policymakers with the option to weigh three competing components when deciding what to prioritize: anticipated effectiveness in \textbf{harm mitigation} through LLM simulations, expert assessed \textbf{cost of exerting a policy}, and a \textbf{participatory element} of what lay stakeholders want enacted in policy. We use a genetic algorithm simulation grounded in lay stakeholder-generated and evaluated scenarios to identify policy sets that maximize perceived harm reduction while balancing resource costs and public perception. We demonstrate the method in the context of generative AI harms in the media environment \cite{kieslich2025anticipating}, specifically \textit{political manipulation}, \textit{unemployment}, and \textit{media sensationalism}.

We balance expert assessment with a participatory element and consequently show how different knowledge sources complement each other in identifying and prioritizing policies. We demonstrate how the method can be adapted to inform policy assessment practices by varying the weights assigned to harm mitigation, cost, and participatory components across different runs and impact types using our genetic algorithm. Different balances result in substantially different outcomes; e.g., when all are balanced equally only a few highly rated and low cost policy options are implemented, whereas prioritizing harm mitigation results in a diverse and numerous selection of policy options, while focusing on cost and participatory elements results in more selective coverage of high impact items. These results highlight the importance of balancing expert and lay opinion in tandem with technical approaches to anticipate harm mitigation.

\section{Related Work}

\subsection{AI Governance}

AI governance has evolved into a broad field, incorporating a variety of approaches to address the challenges posed by AI to society \cite{ai2024artificial, ezeani2021survey, roberts2023governing}. 
Prescriptive regulatory approaches such as the EU AI Act \cite{eu_ai_act_52} remain silent about what adequate risk mitigation measures are, 
which gives power 
to providers of regulated AI systems and challenges effective supervision by regulators \cite{orwat2024normative, kieslich2026scenario}. 
Implementing governance mechanisms is further complicated by the multitude of potential risks and the corresponding mitigation approaches, 
though scholars have developed approaches on how to map future impacts and mitigation measures \cite{barnett2025envisioning, barnett2024simulating, kieslich2026scenario, messmer2023auditing}. 
The myriad of potential policy options 
to address AI harms \cite{barnett2025envisioning} poses a significant challenge for regulators who must prioritize policy actions while most resources are scarce \cite{messmer2023auditing}. 

Consequently, this identification and prioritization process is of pivotal importance in the algorithmic governance process. However, research has shown risk prioritization processes are heavily dependent on human interpretation, as it is difficult to quantify the harm caused by risks such as human rights violations \cite{schmitz2025global}. This becomes even more critical when most risk assessments are carried out by technology companies themselves or dedicated third-party auditors, largely ignoring public input 
\cite{orwat2024normative, gillespie2024generative}. The lack of democratic accountability is concerning, given studies have shown stakeholder groups differ significantly in their identification and rating of risks \cite{hartmann2025addressing, reuel2025evaluates}. 
Engaging in participatory AI governance is of pivotal importance to offer commonly underrepresented stakeholders the chance to contribute to technology deployment and risk management \cite{nikolova2014rise, delgado2023participatory}. 

One way to foreground citizens' voices in technology development and risk management is to utilize scenarios \cite{barnett2025scenarios}. Scenarios are narrative artifacts (e.g., short stories) 
that describe plausible future situations grounded in real-world development, policies, and trends \cite{amer2013review, borjeson2006scenario, Hohendanner2024Metaverse, carroll1995scenario}. They help mapping out potential future developments 
\cite{nanayakkara2020anticipatory, varum2010directions} and 
can be used to evaluate the effectiveness of policy proposals \cite{kieslich2025anticipating, barnett2024simulating}. Further, scenarios are particularly apt for participatory approaches aiming to include civil society as they enable citizens to express themselves and to ground their prospections of the future in their own lived reality \cite{nikolova2014rise, burnam2015creating}. In this study, we use lay stakeholder evaluated scenarios and policy measures, and scale up the evaluation of 
scenarios and mitigation options by using LLMs to support this task. 

\vspace{-3mm}
\subsection{AI for Policy Making}

Regulators 
have the critical task to not only identify and rate the risks of AI systems, but also to embed multiple stakeholder perspectives in the process and do so with often limited resources. One way to address this challenge is to use precisely those data-driven technologies 
for policy-making \cite{rieder2016datatrust, starke2020artificial, verhulst2019data, poel2018big}. AI can influence the whole policy cycle, starting with policy planning to support framing problems for policymakers \cite{hollywood2025well_tempered_ai}, policy formulation to help policymakers sift through different policy options \cite{kuo2025policycraft, valle2020assessing, saxena2024ai, hill2025pen}, policy implementation to automate routine tasks \cite{valle2020assessing}, and policy assessment to support adjusting policies \cite{valle2020assessing, lim2025ai}. 
For all stages, ``simulations allow better solutions to be given or certain policies to be discarded from the agenda because they do not obtain the expected results'' \cite{valle2020assessing}. 
Generative AI can support these practices by simulating future scenarios with realistic risk and policy elements \cite{barnett2024simulating, buccinca2023aha, coz2025would, ferrer2025time}.


With the emergence of large language models (LLMs), scholars have begun using them as a tool to model human behavior at scale \cite{binz2024centaur, anthis2025llm, hu2025simbench, park2023generative, park2024generative}. 
This method has been applied in 
public health for vaccine hesitancy \cite{hou2025can}, economics to simulate studies prior to launching human trials \cite{horton2023large}, and to support policymaking \cite{Richter_Campbell_Riedl_2025}. LLMs 
can also help develop reliable evaluations comparable to crowdsourced human evaluations \cite{dong2024can}. These models 
can sometimes predict treatment effects better than human forecasters \cite{hewitt2024predicting}, showing the promise of LLMs augmenting traditional experimental methods. 

LLMs are well-suited for averages, norms, and treatment effects, though less suited for identifying specific and context-dependent group behaviors \cite{bisbee2024synthetic, dominguez2024questioning,santurkar2023whose}, 
especially when focusing on marginalized demographic groups \cite{hu2025simbench, wang2025large}. Many weaknesses have been identified with this method in that application, such as bias and uniformity \cite{kozlowski2024simulating}, susceptibility to caricature \cite{cheng2023compost}, and generalization \cite{anthis2025llm}. 
Consequently, many researchers call for the importance of validating these simulations against human subjects \cite{kozlowski2024simulating} and have the impacted stakeholders at the forefront of these experiments 
\cite{olteanu2025ai}. Human validation remains an indispensable element to achieve robust and plausible insights into human behavior. Despite these limitations, 
tools. 
LLMs can simulate human behavior in structured, aggregate contexts to support exploratory research, while requiring careful validation and mindful deployment. 
Utilizing generative AI can help 
in the policy prioritization process as it offers the possibility to efficiently simulate multiple policy options. 
In acknowledgment of the risks and technical limitations, we highlight 
the critical role of data selection that should reflect citizen and domain expert input to purposefully balance policy options. We emphasize that AI simulations for policy evaluations should support, never replace, decision-makers in early policy planning phases.



\vspace{-1mm}
\section{Methodology}

\begin{figure}
    \centering
    \includegraphics[width=0.99\linewidth]{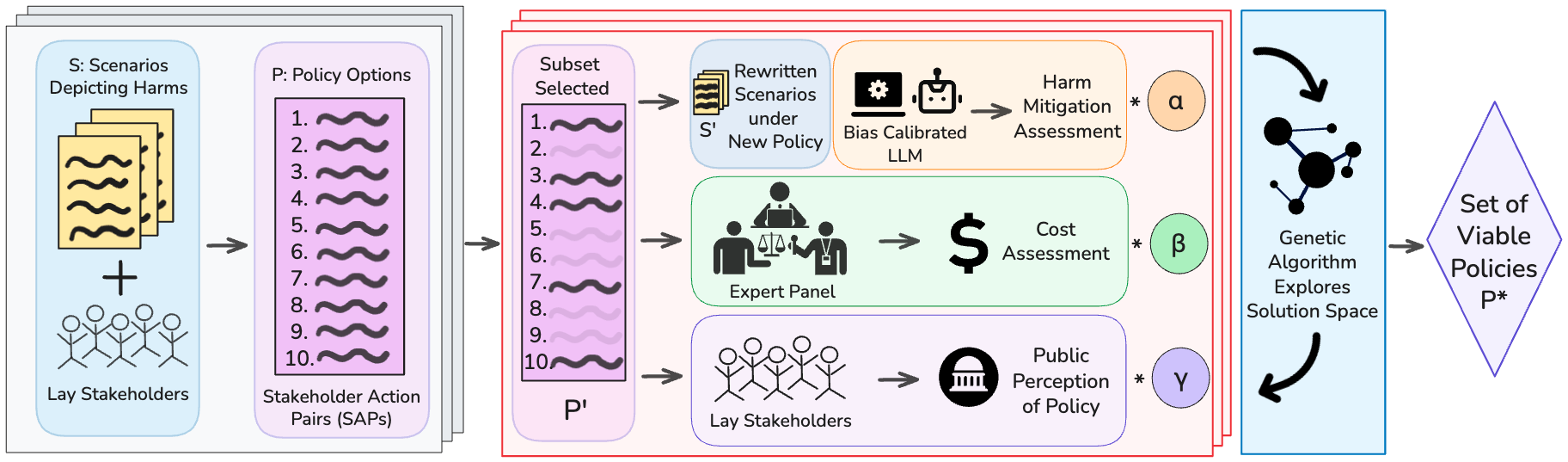}
    \caption{Illustration of our methodology. Scenarios depicting harms ($S$) are evaluated by lay stakeholders and used to brainstorm potential mitigation strategies ($P$) as stakeholder action pairs (SAPs). Each element in $P$ is assigned a cost (C) by an expert panel and evaluated by lay stakeholders (D). A subset of $P$ ($P'$) is selected to rewrite the original scenarios. This is evaluated by an LLM aligned with lay stakeholder evaluators for perceived harm mitigation (weighted by $\alpha$; Sec. \ref{sec:alpha}), and $P'$ is assessed for cost (weighted by $\beta$; Sec. \ref{sec:expert_cost_method}) and public perception (weighted by $\gamma$; Sec. \ref{sec:gamma}). This happens many times as a genetic algorithm explores the solution space $P$ to find a set of viable policies ($P*$) to optimize across these three dimensions.
    \vspace{-5mm}}
    \label{fig:hero_figure}
\end{figure}

At a high-level (Figure \ref{fig:hero_figure}) we propose a method using LLMs to write and rewrite scenarios describing impacts of AI under different policy conditions, evaluating policies by using carefully prompted LLMs to assess the severity, magnitude, and plausibility of impacts as expressed in the scenarios. We employ a genetic algorithm to explore various locally optimal\footnote{We stress that we refer to ``optima'' in a mathematical sense and not in a normative sense; that is, the genetic algorithm calculates local optima, yet the effectiveness and viability of recommended policy options should be interpreted by policymakers not LLMs.} combinations of options which might be stipulated in policy to mitigate specified harms. We choose a genetic algorithm for its efficacy at exploring large optimization spaces as our solution space is intractably large; for \textit{political manipulation} alone, the possible policy options are $2^{31}$, or just over 2 billion (see Sec. \ref{sec:genetic_alg}). We focus on exploring several highly viable local optima since computing the global optimum would be prohibitively expensive (computationally, financially, and ecologically) for real-world policy implementation consisting of many harms and mitigating actions. 
The following sections outline the optimization function, scenario assessment and alignment with human ratings, and the genetic algorithm and datasets we use. 

\subsection{Optimization Function}

 We define a function to guide the optimization across three separately weighted and competing factors: (1) $\alpha$: the weight assigned to the amount of anticipated \textbf{harm mitigation} (\textit{M}) as perceived by lay stakeholders as a function of severity (\textit{Sev}) and magnitude (\textit{Mag}) of impact, (2) $\beta$: the weight assigned to presumed \textbf{cost} (\textit{C}) to implementing the mitigating action as allocated by a panel of experts, and (3) $\gamma$: the weight assigned to the \textbf{participatory} (``democratic'') component (\textit{D})---the prioritization and belief of lay stakeholders that these actions should be pursued. We set it up with these three components weighted separately in order to experiment with how variations in weights affect identified optimal policies. 
 The function takes as an input a set of scenarios \textit{S} that depict a given harm (e.g., \textit{political manipulation}), and a subset $P'$ which reflects a set of interventions or actions to be taken that might be stipulated in a policy designed to mitigate that specified harm. We further define a rewritten scenario $S'$ in which the policy implied by $P'$ was deployed and the harm mitigated to some degree as expressed in the rewritten scenario. A simplified form of this optimization function can be seen in Equation \ref{equ:short_optimization} below, which is further explained more explicitly in the following subsections:


\vspace{-4mm}
\begin{equation}
F(S,P') = \alpha(M) -\beta(C) + \gamma(D)
\label{equ:short_optimization}
\end{equation}

In this work we focus on three harms resulting from the use of generative AI in the media ecosystem: \textit{political manipulation}, \textit{unemployment}, and \textit{media sensationalism}. These harms are selected from a taxonomy of 50 proposed by Kieslich et al. \cite{kieslich2025anticipating}, and in the top five most severe of this taxonomy as evaluated by lay stakeholders \cite{barnett2024simulating, barnett2025envisioning}. 

\subsubsection{Inputs: \textit{S} and $P'$}

Each input to the function includes a set of short narrative scenarios, (\textit{S}),  depicting the specified harm about five years in the future in the United States. For each of the harms we simulate, we use a set of three scenarios depicting that harm, i.e., $||S||=3$ and $S=\{S_1, S_2, S_3\}$. These were generated using GPT-4 in 2024 in our prior work \cite{barnett2024simulating}, and have been validated by our author team for validity and presence of specified harms. The harms depicted in these scenarios have also been evaluated by lay stakeholders for 
severity, magnitude, and plausibility; for more information on this evaluation, please consult our prior work \cite{barnett2024simulating}. The average scenario length is 342 words, and  
all input scenarios ($S$) along with some example rewritten scenarios ($S'$) can be found in Appendix \ref{sec:example_scenarios}. 

We also pass to the function a set $P'$ depicting which options should be stipulated in policy designed to mitigate the specified harm; $P' \in P$, where $P$ is the set of all mitigating actions that might be stipulated in policy for a specific impact type. This set $P$ was brainstormed by lay stakeholders in response to the same stimulus scenarios ($S$) to mitigate the harms depicted in those scenarios \cite{barnett2025envisioning}. We term these policy options as ``stakeholder-action pairs'' (SAPs) in line with \cite{barnett2025envisioning} due to their formative nature of an action taken by a specified stakeholder (e.g., ``News publishers should label content generated with the assistance of AI''). We consolidated the SAPs in \cite{barnett2025envisioning} such that any that were too similar were collapsed under one; we further assessed incompatibility and aside from a couple SAPs banning aspects of AI, there were few incompatible pairs. Even when similar, each policy item remains an independent action. There were 31 possible SAPs in $P$ for \textit{political manipulation}, 16 for \textit{unemployment}, and 26 for \textit{sensationalism}.  Each SAP was brainstormed by lay stakeholders; 
an example list can be found in Table \ref{tab:example_SAPs}, with a full list of possible SAPs in Appendix \ref{sec:appendix_full_sap_tables} Tables \ref{tab:Po_Ma_results}-\ref{tab:MQ_Se_results}. The set of possible policies that could be enacted for each impact type is $2^n$, where $n$ represents the number of SAPs enumerated for the impact type; for \textit{political manipulation} this is $2^{31}$, or just over 2 billion.

\begin{table*}[t]
\addtolength{\tabcolsep}{-0.05em}
\begin{tabular}{c|l|c|ccc}\toprule
 \multicolumn{6}{c}{\textbf{Example Stakeholder Action Pairs (SAPs) for Political Manipulation}} \\
 \cmidrule(lr){1-6}
 & & \textbf{Cost (C)} &
 \multicolumn{3}{c}{\textbf{Participatory Rating (D)}} \\
 & & Panel Average &
 Prior. & Agr. & Score\\
 \multicolumn{1}{c|}{\textbf{Stakeholder}} &
 \multicolumn{1}{c|}{\textbf{Action}} &
 (1-3 $\uparrow$; 4 = NV) &
 (1-3 $\uparrow$) &
 (1-7 $\uparrow$) &
 (Pr. * Ag.)\\
 \cmidrule(lr){1-1}
 \cmidrule(lr){2-2}
 \cmidrule(lr){3-3}
 \cmidrule(lr){4-6}
 News publishers &
 be transparent about any use of AI in stories. &
 1 &
 2.82 & 7 & 19.74\\
\cmidrule(lr){1-6}
Social med. comp.  & 
give users more control over newsfeeds. &
1.75 & 
2.36 & 6.21 & 14.66\\
\cmidrule(lr){1-6}
Schools	&
strengthen digital literacy \& crit. engagement. &
3 & 
2.75 & 6.56 & 18.04 \\
\cmidrule(lr){1-6}
\multirow{2}{*}{Government} &
take legal action against news organizations &
\multirow{2}{*}{4} & 
\multirow{2}{*}{2.4} & \multirow{2}{*}{5.8} & \multirow{2}{*}{13.92}\\
 & 
spreading fake news. &
 & 
 &  & \\
\bottomrule
\end{tabular}
\caption{Four example stakeholder action pairs (SAPs) for \textit{political manipulation}. Includes expert assessed cost (score reported is the average cost from the panel); if an SAP was rated as nonviable (NV) by any expert it was given a 4 (see Sec. \ref{sec:expert_cost_method}). Also includes the participatory rating comprised of lay stakeholder evaluated priority (ternary scale) and agreement (Likert scale 1-7) and the final score which is a product of the priority and agreement. 
All 31 SAPs for this impact are in Appendix Table \ref{tab:Po_Ma_results}.
\vspace{-6mm}} 
\label{tab:example_SAPs}
\end{table*}

\subsubsection{Harm Mitigation $\alpha(M)$}\label{sec:alpha}

This portion of the optimization function $F(S,P')$ represents the degree of anticipated harm mitigation implementing the various SAPs stipulated in $P'$ would achieve. Our data is aligned with \textit{perceived} harm mitigation, not \textit{actual} harm mitigation. Our method orients resources towards promising directions where policymakers would then use existing policy evaluation methods  to measure harm mitigation more holistically; our method is complementary and includes perspectives commonly neglected. We use the same evaluation detailed in prior work \cite{barnett2024simulating} in which we used an LLM to generate scenarios depicting harms and then rewrite them given a policy condition. Lay stakeholders then evaluate the pre- ($S$) and post ($S'$) policy scenarios across independent risk assessment dimensions including severity, magnitude, and plausibility, each evaluated on a scale of 1-5 (Severity: (1) ``Not Severe'' to (5) ``Extremely Severe'', Magnitude: (1) ``A small number of people'' to (5) ``The majority of people in society'' (5), and Plausibility: (1) ``Not plausible'' to (5) ``Extremely plausible''). 

We use a set of SAPs ($P'$) to represent the policy condition under which the scenario $S_k \in S$ should be rewritten, resulting in $S'_k$. We then evaluate the change in severity score ($Sev(S_k) - Sev(S'_k)$) and the change in magnitude score ($Mag(S_k) - Mag(S'_k)$). If the plausibility score of the scenario $S_k$ is below a threshold ($<3$ on a 5-point scale), we remove $P'$ from evaluation, though this rarely occurred ($<0.1\%$). In \cite{barnett2024simulating} each scenario was evaluated by lay stakeholders, however in order to explore the vast policy solution space we instead use a model aligned to scores assigned by lay stakeholders to generate the severity, magnitude, and plausibility scores (see Sec. \ref{sec:model_alignment}). 

We rewrite each scenario $S_k$ in $S$ under the same policy condition $P'$, and calculate the change in severity and magnitude. We take a weighted sum of the change in severity score weighted by $w_s$ and the change in magnitude weighted by $w_m$. Policy analysts using this method could choose these weights according to their own values, however, in the results presented in this paper, we used $w_s=0.65$ and $w_m=0.35$ giving greater weight to severity in line with other work assessing the higher relevance of severity for risk assessment prioritization \cite{gellert2020risk, messmer2023auditing, barnett2025envisioning}. These weighted sums are then averaged over each scenario in $S$. This component is detailed mathematically as:

\vspace{-3mm}
\begin{equation}
\alpha(M)
 = \alpha \left[\frac{1}{||S||}\sum_{k \in ||S||}\begin{aligned}
& w_s\bigl(\mathrm{Sev}(S_k)-\mathrm{Sev}(S_k')\bigr) {}+ \\
&  w_m\bigl(\mathrm{Mag}(S_k)-\mathrm{Mag}(S_k')\bigr)
\end{aligned}
\right]
\end{equation}


\subsubsection{Expert Cost $\beta(C)$}\label{sec:expert_cost_method}

This portion of $F(S,P')$ represents the expert weighted cost assessment, designated by the set $C$. The cost is fixed and only corresponds to the input SAP set $P'$; it is unrelated to the corresponding scenarios $S$. To assign values to each SAP, we had all members of the interdisciplinary author team independently evaluate each SAP for the estimated cost burden that would be placed upon the stakeholder assigned the action (average pairwise Cohen's Kappa of $0.481$, indicating moderate agreement). Members of this author panel include experts in policy, law, journalism, and deployment of AI systems. We assign the expected cost value on a 1-3 scale. 1 is for ``Low Cost'': manageable within existing budgets. 2 is for ``Moderate Cost'': requires some reallocation or new funding. 3 is for ``High Cost'': requires substantial new funding or major operational change. We also had the option to rate the SAP as 4 for ``Nonviable'' (NV): legally or technically infeasible. This nonviable rating could occur when implementing this SAP in a democratic political system is not possible (e.g., government punishing the spread of fake news would impede on freedom of speech and press) or something is technically infeasible for a stakeholder to execute (e.g., technology companies ban usage of AI by news publishers to generate stories). The final cost value was then averaged across the independent evaluators, and if any individual assigned a 4, the final cost was a 4 due to the domain expert judgment overruling the other evaluations. 

$C$ then contains the cost value assigned to each SAP in $P'$. In the event that the SAPs stipulated in $P'$ included any with a cost value of 4, 
those were excluded from evaluation to avoid simulating nonviable policy options. By allowing selection of nonviable pairs if cost is excluded (i.e., $\beta = 0$), we are able to demonstrate in our results the importance of including expert assessment. However, in practice we recommend nonviable SAPs should simply be excluded, reflecting a hard constraint not folded into the cost.
The final cost value assigned to $P'$ is a sum of the costs $C_j$ 
such that $P'$ contains no SAPs with an infeasible cost value assigned, where $n$ is the number of stipulated SAPs. We are optimizing for minimum cost, so this value is negative in $F(S,P')$:
\vspace{-2mm}
\begin{equation}
-\beta(C) = 
 -\beta \left[\sum_{j=1}^{n}  (C_j*P'_{j})\right]
\end{equation}

\subsubsection{Participatory Rating $\gamma(D)$}\label{sec:gamma}

This portion of the optimization function $F(S, P')$ corresponds to the participatory or ``democratic'' component, designated by the set $D$. Similar to the cost $C$, the participatory value $D$ is fixed and only corresponds to the input SAP set $P'$ and has nothing to do with the corresponding scenarios $S$. This component represents the degree to which lay stakeholders want to see these SAPs enacted in policy. These scores were compiled in prior work \cite{barnett2025envisioning}, and were captured using the same scenarios depicting the harms in $S$ as stimuli to gauge lay stakeholder value ranking. 
Participants first read the stimulus scenario, then evaluated a set of SAP statements for agreement on a Likert scale of 1 (Strongly Disagree) to 7 (Strongly Agree) and priority on a ternary scale of low, medium, high. We chose Likert scales due their simplicity and well-understood nature in survey design, but there are some resulting limitations such as a coarse scale not being able to capture the nuance of human decision making. Some other noted limitations of these scales (such as central tendency bias) were not empirically evident in our data; respondents were more likely to choose more extreme values. The final value we assign in $D$ is the product of the mean priority score (range [1,3]) and mean agreement score (range [1,7]) to give equal weight to both terms (range [1,21]). $D$ then contains the final participatory value assigned to each SAP in  $P'$. 
We then take the average participatory value to represent the average agreement and prioritization across the SAPs stipulated in $P'$. We chose not to sum these so as not to reward selection of every SAP, but rather a curated selection of SAPs. The final participatory value assigned to $P'$ is: 

\begin{equation}
\gamma(D) = 
\gamma \left[\frac{1}{n}\sum_{j=1}^{n} (D_j*P'_{j})\right]
\end{equation}

\subsubsection{Output (P*)}

The final score outputted by $F(S,P')$ represents a weighted balance of maximizing the harm mitigation component ($M$), minimizing the expert cost component ($C$), and maximizing the average participatory component ($D$). The data was normally distributed, so we use z-score normalization to rescale each variable (severity, magnitude, cost, and participatory score) to have $\mu=0$ and $\sigma=1$ so they can be treated equally in the final evaluation. For each individual variable, we conduct an initial run for parameter estimation in which we randomly generate 1000 unique sets $P' \in P$ and store the mean and standard deviation of the four variables for each impact type separately for normalization purposes. Because each independent component is normalized at $\mu=0$, the resulting output $P^*$ is also normally distributed with a positive skew ($\mu=0.63$, $\sigma=1.01$) because two components are added and one is subtracted. Assuming all components are weighted equally, a high $P^*$ indicates this policy optimized for high harm mitigation, high participatory element, and low cost; a low $P^*$ means the opposite is true. A neutral $P^*$ likely means two competing factors balance each other out (e.g., high harm mitigation but also high cost), or a modest score across all three. 
The expanded form of the optimization function detailed above (Equation \ref{equ:short_optimization}) with all components detailed as described in this section can be seen below:

\begin{equation}
F(S,P') = \alpha \left[\frac{1}{||S||}\sum_{k \in ||S||}\begin{aligned}
& w_s\bigl(\mathrm{Sev}(S_k)-\mathrm{Sev}(S_k')\bigr) {}+ \\
&  w_m\bigl(\mathrm{Mag}(S_k)-\mathrm{Mag}(S_k')\bigr)
\end{aligned}
\right]
-\beta \left[\sum_{j=1}^{n}  (C_j*P'_{j})\right] + \gamma \left[\frac{1}{n}\sum_{j=1}^{n} (D_j*P'_{j})\right]
\label{equ:full_optimization}
\end{equation}

\subsection{LLM Alignment with Human Ratings}\label{sec:model_alignment}

The harm mitigation component of this function ($M$ weighted by $\alpha$) relies on rating scenarios $S$ across different risk assessment dimensions: severity, magnitude, and plausibility.\footnote{This is separate from the participatory component weighted by $\gamma$ in which lay stakeholders evaluate SAPs for agreement and priority.} The ideal would be to have lay stakeholders evaluate each scenario, but that would not be possible from a cost or time perspective for the scale of simulation we undertake. As a proxy, we align an LLM with human evaluators to evoke what humans might rate each scenario. The dataset we use for this task is lay stakeholder evaluated scenarios created in prior work \cite{barnett2024simulating}, which comprises 234 total scenarios across 39 impact types with about 6-7 lay stakeholders evaluating each scenario across each dimension. 

The human raters had high variance in their ratings, with a low average inter-rater reliability score of 0.07 (range [0.03, 0.19]), reflecting the subjectivity of these ratings. Though this entropy is to be expected in a small set of human evaluations, this proved a challenging dataset on which to align a model emulating human evaluations due to subjectivity and variance in rating, so we collect a larger dataset with more robust ratings for the three impact types we focus on in this study (6 scenarios each): \textit{political manipulation}, \textit{unemployment}, and \textit{media sensationalism}. The goal of collecting and analyzing this sample was two-pronged: to understand (a) if there were any systematic differences across demographic groups in relation to how they evaluated these harms, and (b) if our model was aligned with any demographic more than others. Assessing these potential biases is a critical step towards reliably using an LLM to rate policies in the context of our larger simulation. We recruited 109 participants residing in the US through Prolific to provide ratings.\footnote{Each participant evaluated 6 scenarios. The median completion time was 11 minutes to complete this task, and we paid \$3 per participant resulting in an estimated hourly wage of \$$16.36$/hour. We filtered out 6 participants (of the original 115) who failed attention checks.} In order to understand any differences across primary demographic groups, the dataset was created to be representative of self-reported participant gender, race, and age; we had a 50/50 split for female/male, 50/50 white/POC with 12\% Black, 12\% Asian, 14\% Mixed, and 12\% Other, and a median age of 36. 
There were no statistically significant differences among any demographic splits across multiple impact types, and further our ratings model did not significantly align more closely with any demographic group than another. This analysis both confirmed our model is not aligned more closely with any of the demographic groups we assessed over another, nor are there any major differences among groups. 

We use these two datasets (the breadth-focused 234 evaluated scenarios and the depth-focused 18 scenarios) to align an LLM with human ratings of scenarios across three dimensions: severity, magnitude, and plausibility. We use the breadth-focused dataset as a combination of training and validation sets (80/20 split), where the training set is used for prompt iteration and refinement, and use the more robust depth-focused dataset as a test set. Performance on the test set is most relevant here because the data maps to the 3 impact types which we later simulate. During this alignment stage we optimize for correlation between the model outputs and human ratings. 
We explored multiple models and went through many stages of prompt engineering, leveraging insights from \cite{khattab2023dspy} and \cite{caswell2024ai}. Our best overall scoring model was \modelname{o3-mini} with low reasoning effort, with a test set Pearson correlation coefficient of 0.794 in severity, 0.707 in magnitude, and 0.616 in plausibility. LLMs have demonstrated self preference in some ratings contexts \cite{xu2025ai} but if our employed model were biased in this way, this would manifest as inflated deltas for harm mitigation which would not affect the overall optimization since scores are relative. We ran a comparison using \modelname{Claude Opus 4.6} and \modelname{Sonnet 4.5}; though raw scores for original and rewritten scenarios were significantly different at $p<0.05$ (\modelname{Claude} scores were consistently lower than \modelname{o3-mini}), the deltas were not ($p>0.05$). More details on model exploration, full prompts and prompt engineering, and robustness and stress tests can be found in Appendix \ref{sec:evaluating_scenarios_appendix}. 


\subsection{Genetic Algorithm}\label{sec:genetic_alg}

The set of possible policies that could be enacted for each impact type is $2^n$, where $n$ represents the number of SAPs enumerated for the impact type. For \textit{political manipulation} alone, this is $2^{31}$, or just over 2 billion possible combinations. The solution space is intractably large, especially as we are generating and evaluating a set of scenarios for each combination. In order to efficiently evaluate this space, we deploy a genetic algorithm \cite{mitchell1998introduction, rawlins1991foundations, holland1992adaptation}, which uses evolution as an inspiration for exploring complex solution spaces of optimization problems. Due to the manner in which genetic algorithms explore solution spaces, it is not guaranteed to find a global optimum \cite{mitchell1998introduction, wright1931evolution}, and will find any of several possible local optima, though parameterization increases the likelihood of finding the global optimum \cite{de1975analysis, grefenstette1986optimization, schaffer1989study}. 

There are a variety of models that could be employed for an optimization problem such as ours. Beyond the genetic algorithm we employ \cite{mitchell1998introduction, rawlins1991foundations, holland1992adaptation} the most obvious alternative is a family of models called multiobjective evolutionary algorithms (MOEAs) \cite{fonseca1993genetic, Deb_MOEA_2014, zitzler2001spea2, coello1999comprehensive} and most notably among them the Non-dominated Sorting Genetic Algorithm II (NSGA-II) \cite{Deb_NSGA_II}. These optimization algorithms seek to explore the Pareto frontier in which no feasible solution can improve upon another without worsening one criterion (e.g., cost) in order to increase at least one other criterion (e.g., perceived harm mitigation). The core difference between our approach and multi-objective methods like NSGA-II is that rather than discovering the Pareto frontier and choosing among solutions post hoc, we force tradeoffs to be made explicitly up front, empowering users to optimize according to their explicitly-weighted objectives (e.g., prioritize expert opinion over participatory valuation),
ensuring transparency and accountability of stated objectives rather than post hoc choice among alternatives. Other options beyond MOEAs could also be utilized, such as portfolio decision analysis \cite{martin1955mathematical, chang2009portfolio, samuelson1975lifetime}, which similarly results in a portfolio of options and allows exploring tradeoffs post hoc rather than upfront to understand what resulting ``optimal'' options under stated conditions would be. Another comparable method is the knapsack resource allocation problem \cite{salkin1975knapsack}, which selects subsets of predefined items under capacity constraints; our method goes beyond this to optimize a multi-criteria value function (not simply resource constraints) using evolutionary search as an optimization technique. Our approach can then be characterized as a form of multiple-criteria decision analysis \cite{greco2016multiple} utilizing z-score scalarization to define the objective function, then optimized via a genetic algorithm.\footnote{We make our source  code available at the anonymous Git repo: \url{https://github.com/informingaipolicyassessment/InformingPolicyAssessment}.}

Genetic algorithms typically function by first generating a random initial population of ``chromosomes;'' in our case each chromosome is a different subset of SAPs $P'$. Each chromosome 
is then evaluated according to some fitness function, in our case $F(S,P')$ (Equation \ref{equ:short_optimization}). The next generation of the population is created by selecting parent chromosomes ($P^{'a}$ and $P^{'b}$) to crossover to produce new offspring ($P^{'c}$ and $P^{'d}$). There is then a chance that the offspring are randomly mutated before entering the next generation's population. This selection, crossover, mutation process is repeated until the population reaches a desired size. Then each member of the next generation is evaluated, and the cycle repeats until a stopping criteria is met. The criteria we use for stopping is 
maximum number of stall generations reached in which no new optima is found. We go into each stage more in depth below; we provide high level information about parameter selection in this section with additional details in the Appendix \ref{sec:appendix_gen_alg_details}.

\subsubsection{Population Initialization}

Based on standard settings from \cite{de1990analysis, grefenstette1986optimization} and scaling recommendations from \cite{goldberg1991genetic,carroll1996genetic}, we choose population sizes based on the following formula: $Pop Size = \frac{||P'||*2^{\bar{P'}}}{\bar{P'}}/2$, where $||P'||$ is the length of $P'$, and $\bar{P'}$ is the average number of nonzero elements in $P'$. For example, there are 22 feasible SAPs for political manipulation, and, based on an initial parameter estimation run, $P'$ had on average 7 nonzero encoded elements. So the default population size becomes $\frac{22*2^{7}}{7}/2=201$. For \textit{unemployment} and \textit{sensationalism}, we use 105 and 136, respectively. We initialize a population of random unique chromosomes (i.e., subsets $P'$). We use our fitness function $F(S,P')$ to evaluate each chromosome, and these evaluated chromosomes become the base for the first generation of our genetic algorithm.

\subsubsection{Parent Chromosome Selection}

To create each subsequent generation, we select two parent chromosomes from the previous generation to create two offspring for the next generation. The original genetic algorithm implemented by Holland implements fitness-proportionate selection \cite{holland1992adaptation}, which means the chance of being selected to be a parent corresponds to how well a chromosome scored via the fitness function $F(S,P')$. The most common method for implementing this fitness-proportionate selection is called ``roulette-wheel sampling'' \cite{golberg1989genetic}, which can be thought of as assigning each chromosome a wedge of the roulette wheel proportional to how well it scored on the fitness function relative to all other possible parent chromosomes. This means every chromosome has a chance, though those that scored well on the fitness function have a much higher chance, of being selected. 

\subsubsection{Crossover, Mutation, and Elitism}

Once we have two parents selected, we now have to create two offspring chromosomes from them. The main parameters that matter here are: (1) crossover chance \& crossover type, and (2) mutation chance. Crossover chance corresponds to the chance that the two selected parents crossover to create new offspring, or if they are simply added to the next generation as is (without crossing over). Literature suggests a high crossover rate to ensure sufficient variation to approach the optimal solution more rapidly and avoid local optima \cite{grefenstette1986optimization, carroll1996genetic, de1990analysis, mitchell1998introduction}, so we implement two-point crossover with a  crossover chance of 0.80. Mutation chance means just before two chromosomes are about to enter the population, each element has a low chance of ``mutating'' to its opposite value (e.g., $1\rightarrow0$ or implementing an SAP or not); we implement a low mutation chance of 0.03. At the end of all of this, there is still a slim chance that the best chromosome (subset $P'$) was not brought forward into the next generation. To avoid this, we implement a concept called ``elitism'' \cite{de1975analysis}, in which the top $k$ chromosomes from the previous generation are brought into the next generation prior to creating any new offspring. This ensures that the best individuals are not lost if not selected for reproduction or whose fitness is reduced by crossover and mutation; we choose $k=3$. 

\section{Results}

\subsection{Overall Results of Optimal Policies for Varying Weights}\label{sec:overall_results}

\begin{figure}[htbp]
    \centering
    \includegraphics[width=0.99\linewidth]{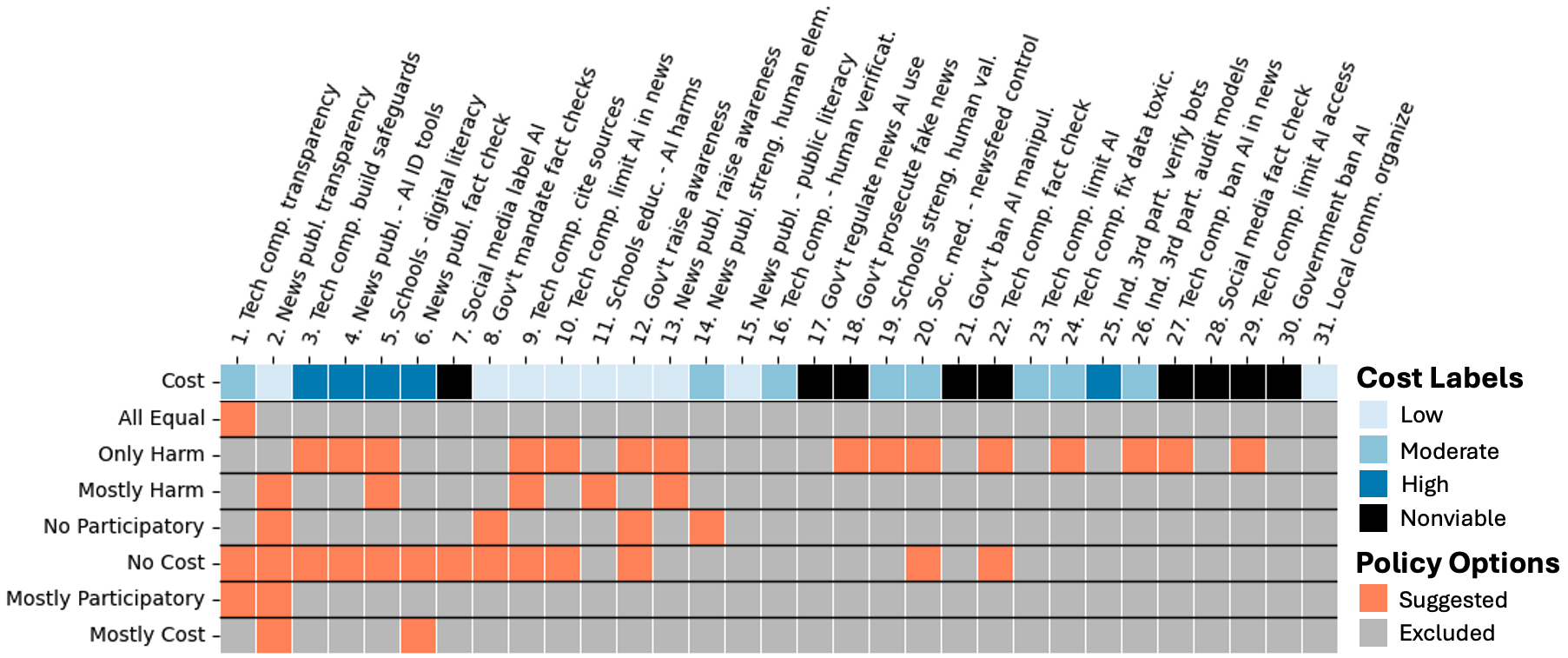}
    \caption{A heat map displaying the final suggested policy for each weight set for \textit{political manipulation}. Shorthand versions of the SAPs correspond to the full description in the Appendix Table \ref{tab:Po_Ma_results}. Orange boxes indicate when an SAP was suggested, gray when it was not. The top row indicates the cost of each SAP with darker color meaning higher cost.}
    \label{fig:heatmap_pol_small}
\end{figure}

This method allows policymakers and researchers to explore various tradeoffs in viable policy options when assigning different weight to potential for harm mitigation, expert cost assessment, and participatory priority. We ran this model with varying sets of weights for each of the three impact types separately, and explored solution spaces when weighting the three components of the equation differently: the harm mitigation component ($M$ weighted by $\alpha$), the expert assessed cost component ($C$ weighted by $\beta$), and the participatory component ($D$ weighted by $\gamma$). Table \ref{tab:descriptive_results} displays the combinations of weights we tried and a description of the optimal policy resulting from those runs, and Figure \ref{fig:heatmap_pol_small} shows which policies were suggested for each set of weights. Figures \ref{fig:pol_heat_map_saps}-\ref{fig:meq_heat_map_saps} in the Appendix provide more detailed results.

\textbf{All Equal}. When all weights are held equal, the suggested policy tends to only specify 1-2 SAPs of low-moderate cost, heavily weighted towards the top of participatory preference. The average change in severity and magnitude is modest compared to all other runs in which harm mitigation ($\alpha$) holds more weight, but overall the optimal policy balanced participatory elements with low costs. In \textit{political manipulation}, this involved only the moderate cost high participatory-rated SAP, ``Technology companies should be transparent about data collection and usage.''

\begin{table*}[htbp]
\addtolength{\tabcolsep}{-0.05em}
\begin{tabular}{ccc|l|cc|ccc|ccc}\toprule
 \multicolumn{12}{c}{\textbf{Overall Results from Varying Weights}} \\
 \cmidrule(lr){1-12}
 \multicolumn{3}{c|}{\textbf{Weight}} &
 \multicolumn{1}{c|}{\multirow{2.5}{*}{\textbf{High Level Description}}} &
 \multicolumn{2}{c|}{\textbf{Avg. $\Delta$}} &
 \multicolumn{3}{c|}{\textbf{Count of SAPs}} &
 \multicolumn{3}{c}{\textbf{\% of SAPs}}
 \\
 \cmidrule(lr){1-3}
 \cmidrule(lr){5-6}
 \cmidrule(lr){7-9}
 \cmidrule(lr){10-12}
 \
 $\alpha$ & $\beta$ & $\gamma$ &
 &
 Sev & Mag &
 Mean & Min. & Max. &
 Mean & Min. & Max.  \\[-0.5pt]
 \cmidrule(lr){1-12}
 0.34 & 0.33& 0.33 &
 Only 1-2 SAPs; highest participatory &
 \multirow{2}{*}{1.28} & \multirow{2}{*}{1.08} &
 \multirow{2}{*}{1} & \multirow{2}{*}{1} &  \multirow{2}{*}{2} & 
 \multirow{2}{*}{6\%} & \multirow{2}{*}{3\%} &  \multirow{2}{*}{13\%} \\ 
 \multicolumn{3}{c|}{\textit{All Equal}} &value and low cost combinations &&&&&&&&\\ [-0.5pt]\cmidrule(lr){1-12}
 1 & 0 & 0 &
 Largest amount of SAPs mandated;&
 \multirow{2}{*}{1.76} & \multirow{2}{*}{1.60} &
 \multirow{2}{*}{11} & \multirow{2}{*}{7} &  \multirow{2}{*}{15} & 
 \multirow{2}{*}{44\%} & \multirow{2}{*}{35\%} &  \multirow{2}{*}{50\%} \\ 
 \multicolumn{3}{c|}{\textit{Only Harm}} &40-50\% of SAPs suggested &&&&&&&&\\ [-0.5pt]\cmidrule(lr){1-12}
 0.5 & 0.25 & 0.25 &
 Almost all low-moderate cost \& &
 \multirow{2}{*}{1.58} & \multirow{2}{*}{1.43} &
 \multirow{2}{*}{5} & \multirow{2}{*}{3} &  \multirow{2}{*}{8} & 
 \multirow{2}{*}{19\%} & \multirow{2}{*}{15\%} &  \multirow{2}{*}{26\%} \\ 
 \multicolumn{3}{c|}{\textit{Mostly Harm}} &in upper echelon of participatory &&&&&&&&\\[-0.5pt] \cmidrule(lr){1-12}
 0.5 & 0.5 & 0 &
 On average 3 SAPs, all low-mod. &
 \multirow{2}{*}{1.67} & \multirow{2}{*}{1.49} &
 \multirow{2}{*}{3} & \multirow{2}{*}{2} &  \multirow{2}{*}{4} &
 \multirow{2}{*}{13\%} & \multirow{2}{*}{8\%} &  \multirow{2}{*}{19\%} \\ 
 \multicolumn{3}{c|}{\textit{No Participatory}} & cost; top of participatory&&&&&&&&\\[-0.5pt] \cmidrule(lr){1-12}
 0.5 & 0 & 0.5 &
 Large number of SAPs, high costs  &
 \multirow{2}{*}{1.73} & \multirow{2}{*}{1.63} &
 \multirow{2}{*}{7} & \multirow{2}{*}{3} &  \multirow{2}{*}{13} & 
 \multirow{2}{*}{26\%} & \multirow{2}{*}{19\%} &  \multirow{2}{*}{42\%} \\ 
 \multicolumn{3}{c|}{\textit{No Cost}} & (nonviable); top half participatory&&&&&&&&\\ [-0.5pt]\cmidrule(lr){1-12}
 0.25 & 0.25 & 0.5 &
 Typically 2 items, included high cost  &
 \multirow{2}{*}{1.19} & \multirow{2}{*}{0.94} &
 \multirow{2}{*}{2} & \multirow{2}{*}{1} &  \multirow{2}{*}{2} & 
 \multirow{2}{*}{7\%} & \multirow{2}{*}{4\%} &  \multirow{2}{*}{13\%} \\ 
 \multicolumn{3}{c|}{\textit{Mostly Participa.}} &SAPs in top of participatory&&&&&&&&\\ [-0.5pt]\cmidrule(lr){1-12}
 0.25 & 0.5 & 0.25 &
 Typically 2 items, moderate cost  &
 \multirow{2}{*}{1.42} & \multirow{2}{*}{1.24} &
 \multirow{2}{*}{2} & \multirow{2}{*}{1} &  \multirow{2}{*}{2} & 
 \multirow{2}{*}{7\%} & \multirow{2}{*}{4\%} &  \multirow{2}{*}{13\%} \\ 
 \multicolumn{3}{c|}{\textit{Mostly Cost}} &SAPs in top of participatory&&&&&&&&\\ [-0.5pt]\cmidrule(lr){1-12}
\bottomrule
\end{tabular}
\caption{Overall results from all runs ($n=14$ per impact type) of the genetic algorithm, averaged across the three impact types. Weight displays how much weight we allocated to harm mitigation ($\alpha$), cost ($\beta$), and participatory ($\gamma$) values. For each weight set, we provide a high level description of what the optimal policies specified, the average decrease in severity and magnitude on a 5-point Likert scale, the average number of SAPs specified by the policy, and average percent of SAPs of the total possible specified.} 
\label{tab:descriptive_results}
\end{table*}

\textbf{Emphasis on Harm Mitigation}. When harm mitigation is the only factor we consider ($\alpha=1$; $\beta,\gamma=0$), the results show that by far the most SAPs are being stipulated in the optimal policy solution in comparison to all other conditions. This condition has one of the highest anticipated change in severity and magnitude, but also the largest cost by a great margin. Most of the SAPs were in the upper half of the participatory evaluation, but relatively average compared to other weights. This weight combination also frequently resulted in nonviable SAPs as determined by expert cost assessors, because cost was not considered as a component. For example in \textit{sensationalism} an optimal set of SAPs as suggested by these weights included 12 of 26 possible SAPs, 4 of which were considered nonviable by experts such as ``The government should make it illegal to use AI to misrepresent the facts.''
When harm mitigation was weighted higher than cost and participatory components but they were still included ($\alpha=0.5; \beta, \gamma = 0.25$), we see similar harm reduction with far fewer SAPs dictated than the ``only harm'' condition with a lower cost and higher participatory element. 


\textbf{No Participatory or No Cost}. When either the participatory component was not considered at all ($\alpha, \beta = 0.5; \gamma=0$) or the cost was not considered at all ($\alpha=0.5; \beta=0; \gamma = 0.5$), there was a relatively similar level of harm mitigation. 
When the participatory element was not considered, there was a limited number of SAPs, but it still stayed towards the top of the participatory weighted elements. However when there was no cost included it resulted in the second highest number of SAPs (second only to exclusively harm mitigation), and often included nonviable SAPs as deemed by cost assessment, such as in \textit{political manipulation}, ``Social media companies should have labels identifying when a post is AI generated,'' which is not technically feasible with society's current ability to identify any AI generated content.

\textbf{Mostly Participatory or Mostly Cost}. When either the participatory component ($\alpha, \beta = 0.25; \gamma=0.5$) or cost ($\alpha=0.25; \beta=0.5; \gamma = 0.25$) was weighted higher than the other two components, we see similar results for optimal policy. There are only 1-2 SAPs stipulated in both cases. The main difference is the higher participatory weighted combination sometimes included a high cost SAP, and the cost weighted combination sometimes included a lower evaluated participatory SAP. Harm mitigation is comparable in both cases.

\subsection{Variation of Optimal Policies}\label{sec:results_subsec2}

As detailed in the methods section (\ref{sec:genetic_alg}), a genetic algorithm is not guaranteed to find the globally optimal solution (i.e., the best possible). It will instead find a locally optimal solution (i.e., close to best for a given area of the solution space), and, if run multiple times, can arrive at different locally optimal solutions. We ran the model multiple times for each set of weights and often found slightly different solutions with comparably high scores for the fitness function. There were consistent trends for different sets of parameter weights (see Section \ref{sec:overall_results}), but there were also interesting differences within and across different sets of weights, which can be seen illustrated in Figure \ref{fig:heatmap_pol_small} and Figures A.\ref{fig:pol_heat_map_saps}-\ref{fig:meq_heat_map_saps}.

\textbf{Popular SAPs.} Within all three impact types were some SAPs that were selected more often than others. For \textit{political manipulation}, the moderate-cost, highest-rated SAP by participatory standards was selected in 69\% of the final policies: ``Tech companies should be transparent about data collection and usage.'' The second highest by participatory standards, though low cost in comparison, was present in 50\% of final optimal policies: ``News publishers should be transparent about any use of AI in stories.'' The most commonly selected SAP in \textit{sensationalism} (in 63\% of final policies) was the fifth highest by participatory standards \textit{and} moderate cost, it was consistently chosen when harm mitigation ($\alpha$) was weighted higher than the other components: ``News publishers should fact check all content and reject any sensationalized stories.'' A similar trend emerged for the top two SAPs in \textit{unemployment} (both in 50\% of policies): moderate cost, 2nd quartile by participatory evaluation, but perceived high impact in harm mitigation: ``News publishers should not replace journalists with AI.'' and ``News publishers should strengthen the human element in journalism.''

\textbf{Variation Among Policies}. Both within the same set of weights and across different sets of weights, variation in suggested policy options was more common than exact duplicates in output policies. Two thirds of all of the identified suggested policy options were unique, and there were six sets of duplicate policies, typically within the same weight set (3/6) and some policies that appeared multiple times in different weight sets. Across all three impact types, the policy identified when harm mitigation was not considered ($\alpha=0$) was duplicated within the weight set and only stipulated one low cost high participatory evaluated SAP; for \textit{unemployment}, this was ``News publishers should employ people to fact check AI.'' This same 1-SAP-policy was also included in the \textit{unemployment} output at least once for three other weights: All equal, Mostly cost ($\beta=0.5; \alpha, \gamma = 0.25$), and Mostly participatory ($\gamma=0.5; \alpha, \beta = 0.25$). Across all impacts, when harm mitigation was weighted higher than at least one other component, this resulted in the most diverse policies.

\textbf{Introduction of Nonviable SAPs.} As detailed in the methods section (\ref{sec:expert_cost_method}), a component of the expert cost assessment involved identifying nonviable SAPs for either political or technical infeasibility. Whenever cost was considered in the algorithm (i.e., $\beta>0$), this meant that these SAPs were removed from consideration. However, when cost was removed from the equation ($\beta=0$), it was possible for the model to select these infeasible SAPs. For the runs in which this was possible, at least one nonviable SAP was chosen 67\% of the time. It was chosen in every run for \textit{sensationalism}, 75\% of the time for \textit{political manipulation}, and 25\% for \textit{unemployment} (though there was only 1 nonviable SAP within \textit{unemployment}). This highlights the importance of including expert assessment. 

\textbf{Duplicate SAPs' Varied Performance.} There were a limited number (10/62) of duplicate SAPs across impact types---those that were brainstormed by lay stakeholders for multiple impact types. There were 7 shared across \textit{political manipulation} and \textit{sensationalism}, 1 across \textit{political manipulation} and \textit{unemployment}, 1 across \textit{unemployment} and \textit{sensationalism}, and 1 across all three. Some of these were nonviable and thus rarely if ever chosen (cost had to be weighted at $\beta=0$ for this to even be a possibility), but some were popular in multiple impact types. For instance the low cost SAP ``News publishers should be transparent about any use of AI in stories'' was present in 50\% of \textit{political manipulation} optimal policies and 56\% of \textit{sensationalism} policies. 
The SAP shared across \textit{manipulation} and \textit{unemployment}, ``News publishers should strengthen the human element in journalism,'' was present in 50\% of \textit{unemployment} optimal policies but only 19\% of \textit{manipulation} policies. The popular SAP shared across all three, ``News publishers should fact check all stories,'' was present in 38\% of \textit{manipulation} runs, 44\% of \textit{unemployment} runs, and 63\% of \textit{sensationalism} runs.

\subsection{Combining Impact Types}

The majority of this work involves treating each impact type independently with separate scenarios and SAPs. In practice, policymakers may need to optimize over several impacts simultaneously. To address this, we explore what happens when we optimize over all SAPs across all three impact types. For this section, we use 2 scenarios per impact type, (6 total) and have each of them rewritten with the same subset of combined SAPs $P'$; harm mitigation is then averaged across all scenarios. Full results in Appendix Table \ref{tab:overlapping_saps} and Figure \ref{fig:massive_combined_heatmap}.  Some SAPs were duplicates across impact types (10/62) as detailed above in Section \ref{sec:results_subsec2}, but the remaining 52 were unique to their impact type. Even though the overlapping SAPs make up a low percentage of the total options and some were nonviable, they still frequently appear in the suggested SAPs identified by the model across different weight conditions.

The most common overlapping SAP that appeared in these combined runs was the high cost ``News publishers should fact check all stories,'' and it was the only one originally present in all three impact types. Because there were so many possible SAPs, the cost of the SAP seemed to matter less than the overall number of SAPs implemented. When cost is not a factor ($\beta=0$), the model on average implemented $23/62$ SAPs, favoring SAPs from \textit{political manipulation} and \textit{sensationalism}. When the participatory element was not a factor ($\gamma=0$), the SAPs were split fairly equally among all three impact types, and it favored a lower amount of SAPs total (average of 11/62). When all components were weighted equally, the model suggested on average $8/62$ SAPs of all costs and all on the upper half of the participatory evaluation. Though 33\% were overlapping impacts, 73\% were from \textit{political manipulation} and only two were from \textit{unemployment}. Some solutions for impacts may be too niche (e.g., involving unions) to address a wider array of harm mitigation.


\section{Discussion}

The introduction of AI in society introduces manifold challenges for policymakers. Many countries introduce risk management approaches to accompany this process to protect societal values while simultaneously not stifling innovation potential \cite{de2022european, van2021risk}. Consequently, AI risk management 
entails deciding on crucial tradeoffs for policy implementation, for instance, weighing the financial and legal cost of a policy proposal against its 
effectiveness to mitigate harms \cite{messmer2023auditing}. 
In practice, regulators and risk assessors have limited guidance on how to decide on these tradeoffs due to a lack of public input in assessing policy options as well as a lack of data and decision support for rating competing elements of the tradeoff \cite{kieslich2026scenario, schmitz2025global}. This paper presents a method that can enable policymakers and researchers to explore viable policy options while incorporating different stakeholder input. 
When designing policy it is essential to consider not only the extent of harm mitigation but also costs, social acceptability, and political feasibility of implementation---factors 
which our proposed method allows policymakers to weigh alongside harm mitigation. We demonstrate that varying the weights assigned to these three components consequentially affects which policy options emerge as viable. 

Based on the use case of harms of the use of generative AI in the media environment, 
we evaluated different policy combinations comparing the different weight options of the factors harm mitigation, 
assessed costs by experts, 
and prioritization by the public. 
The weighing of the different factors leads to largely different policy solutions, differing not only in the amount of anticipated harm mitigation, but also in number and substance of recommended policy suggestions. For instance, when prioritizing harm mitigation over cost and participatory elements, 
our model suggests the enactment of a higher number of policy options. 
As a result, these conditions are also associated with high harm mitigation potential. In contrast, when weighing costs higher, 
fewer policies are recommended. 
These options, while less costly to enact, are perceived as less effective in mitigating harm. Weighing all factors equally leads to the recommendation of 1-2 SAPs with a medium effectiveness for harm reduction, but the highest participatory value, i.e., acting most in the public interest. Further, the recommended SAPs also varied between the different weight options as shown in Fig. \ref{fig:heatmap_pol_small} (and A.\ref{fig:pol_heat_map_saps}-\ref{fig:meq_heat_map_saps}), 
but some SAPs are consistently recommended as viable policy options. 

One potential useful approach for policymakers in practice might be to create stages of input from stakeholders. They might first consult a panel of experts which SAPs are nonviable for political, legal, or technical reasons, then run the genetic algorithm under the ``only harm'' condition. After it produces a set of viable policy options, they would have an expert panel assess the cost of implementation. Finally they could surface those options to their constituencies and have them vote on which measures they would like to see implemented. Alternatively they could explore the different weights from the onset to understand what different tradeoffs exist. 
In practice we suggest that policymakers 
(1) collect participatory input from representative samples of impacted 
stakeholders and (2) include appropriate technical, legal, scientific, and civic experts to assess cost, viability, and validity of proposed policies.

This work highlights the importance of combining participatory data from lay stakeholders with expert data \cite{andersen2021stakeholder}. Expert and public assessments often diverge and, depending on who gets prioritized, can lead to materially different outcomes \cite{kieslich2025anticipating, bonaccorsi2020expert, reuel2025evaluates}. Expert opinion is necessary to rule out 
policies that are technologically infeasible or threaten fundamental rights. 
On the other hand, non-experts provide valuable insights into perceived harm reduction as they are normally affected most (e.g., as passive or active users) with the harms of the technology; thus non-experts bring lived experiences that experts might otherwise overlook. This study identifies and describes a mechanism to combine both perspectives and allow policy makers and researchers to determine how much weight to assign to either side. 
While demonstrated using one method to gather participatory data (scenario building), this could also be adapted to other types of participatory AI data collection methods, such as citizen juries \cite{balaram2018artificial, van2021trading}, joint problem formulation \cite{martin2020participatory}, or co-design of digital or physical artifacts to anticipate and understand harms from AI systems \cite{andres2024understanding, klassen2022run, lu2024contamination}. 



\subsection{Limitations and Future Work} 

We want to underline the computational, ecological, and financial costs of the method, in that it relies heavily on the use of generative models (OpenAI's in the results we presented). 
Each set of policy options required (a) rewriting the input scenario according to those policy options and (b) using an LLM to evaluate that scenario. On average, one run of the model for one impact type and set of weights cost \$60-180, and as this process is not instantaneous; it also took a couple days for each model iteration to run. Though this is still a cost-effective exploratory method prior to a randomized control trial in terms of time and money, it is not a negligible cost. OpenAI models also have their own biases and further model families should be tested. We emphasize this is a proof of concept demonstrating the method and tradeoffs rather than a fully resourced implementation. This also means that we are not arguing the parameter decisions we used are the best possible or only options, rather a functional starting point that we landed on after initial pilot studies and experimentation. Future work can explore different parameterizations and ablation studies, such as greedy \cite{papadimitriou1998combinatorial} or integer programming \cite{wolsey2020integer} baselines, varying weights assigned to severity and magnitude, and different normalization techniques. Additionally, as elaborated in Section \ref{sec:genetic_alg}, using a single objective function may miss non-convex tradeoffs that would be uncovered by using multiobjective functions like NSGA-II. Though we emphasize our choice of model enables researchers and policymakers alike to explore tradeoffs with explicitly weighted objectives rather than providing them with a portfolio of options, future work can explore this Pareto non-dominated set and see where both methods can complement each other. 

This method was tested by researchers as a proof of concept, not policymakers; future work will need to assess the viability of incorporating this approach into policy pipelines in practice. The participatory datasets we used were not for any representative citizen base but rather a crowdsourced body of evaluators. In practice this method will have to have data sourced from the constituency base for whom the policymakers or researchers are designing policy, and the diversity of policy options will then also be dependent on the lay stakeholders consulted. Similarly, we acknowledge participants might suggest more creative or novel ideas based on more complex human-written scenarios (vs. LLM-generated scenarios), but we made the methodological decision to simplify the task to make it more practical to complete for lay stakeholders. In practice, policymakers should assess if there are any significant demographic splits in ratings across constituent groups beyond those assessed in this work, such as more granular splits between racial groups. Future work needs to pilot this method with actual policymakers as well as use a panel of experts to assess costs relevant to the constituency at hand, and ideally involve policymakers and practitioners. Even further we used a coarse cost scale that collapses legal, political, and technical viability into one measure; again this was a proof of concept. Our method allows for this cost to be a more complex and robust measure designated by this expert panel.

We emphasize that the policies suggested by this method would undergo further review by these policymakers and are only a starting point. We recommend in practice adding governance guardrails such as mandatory human verification, documentation of model limitations, and fairness constraints to protect minority preferences. Future work could examine how distribution-sensitive objectives could be incorporated, such as minimizing worst-off harm or ensuring minimum benefits across protected groups, to ensure that those most vulnerable benefit from this approach.

\section{Conclusion}

In this work we propose a method to assist policymakers in prioritizing policy options in response to AI harms. We do so by evaluating tradeoffs between policy options while balancing potential for harm mitigation, expert-assessed cost of implementation, and participatory driven public perception and political viability. We demonstrate how the variability in viable policy options identified by a genetic algorithm could be useful in examining the diversity of options for AI governance. 
This work highlights the importance of balancing expert opinion with that of lay stakeholders, and demonstrates how we can utilize generative AI to assist with identifying viable starting points in the policy process.

\begin{acks}
  The authors acknowledge the National Artificial Intelligence Research Resource (NAIRR) Pilot and the provided research compute for contributing to this research result. We also thank Bryan Pardo for his helpful feedback.
\end{acks}

\section{Author Positionally Statement}

All authors declare that they have no conflicts of interest. We would also like to address the positionality of the authors. This is an interdisciplinary team of researchers with expertise across many domains including law, policy, journalism, and deployment of AI systems. We bring our subset of expertise to the ``expert panel'' employed in the work, and emphasize that this expertise should be comprised by policymakers, regulators and other contextually relevant parties when deployed in practice. Two of the authors identify as female and two as male. In addition, two authors have a European background and two have a US background. All authors identify as white. We therefore acknowledge a bias in the conduct of this research, as our research team does not reflect the experiences of non-white and non-binary people, nor other marginalized groups whose voices we seek to amplify through this work. Finally, we all come from countries with a focus on democratic governance that emphasize participatory decision-making, and this has certainly shaped the way we conducted this project.


\section{Ethical Considerations Statement}

Using generative AI in a high stakes domain such as policy making is a delicate process. We want to emphasize that our proposed method is designed as a \textit{support} tool for policymakers and researchers; it is never intended to replace them. We do not believe this should be an automated process but rather a tool policymakers can use to explore the possible space of policy options. We also want to reiterate that this method would not work if the data is not grounded in lay stakeholder input: another aspect that requires careful human involvement and not automation by LLMs. Nevertheless, we acknowledge that designing a methodology that produces ``optimal'' policy, even when we mean this term in the mathematical sense, can be misconstrued as a dangerous interpretation. We have striven to avoid this interpretation in our presentation of the method and results. To be clear: We are not advocating for a purely AI-based policy assessment. Our results should not be understood as an easy technological fix for policy problems; users should not take our method as the sole approach to solve complex policy assessment problems. We do not support the simulation of public opinion as baseline data, for instance, in only tasking LLMs to propose SAPs or evaluate scenarios without relying on human data. 

We also employed crowdworkers to evaluate our ``depth-focused'' dataset in this work. This study was approved under IRB at the host university of the first author, but we acknowledge this is only the bare minimum in ensuring the crowdworkers are treated with respect and dignity. We put effort into ensuring they were paid a fair wage for their time (including pilot testing among the author team and iterating as the study unfolded). We also made sure there would be no burden to crowdworkers beyond potential discomfort with the harms depicted in the scenarios, and made sure they knew they could leave the study at any time without penalty and still be paid for their time. 

\section{Use of AI statement}

The core of our methodology utilized generative AI in order to rewrite and evaluate scenarios at scale, and we include in our literature review grounds for why we think this is both possible and useful. We believe, and argue, this analysis would be prohibitively expensive to conduct at scale if we did not use generative AI. We emphasize throughout the paper the importance of using this methodology grounded in lay stakeholder and expert assessed data---not LLM generated data. We also include in our limitations some issues with using LLMs for these tasks.

We did not use generative AI to conduct the analysis or draft any content for the paper. One author, whose native language is not English, used DeepL for text editing, but not for text generation. One author used GPT-5 for assistance in formatting LaTeX tables and formulas. The other authors did not use generative AI to write this paper. 

\bibliographystyle{ACM-Reference-Format}
\bibliography{references}

\appendix
\setcounter{secnumdepth}{2}
\section{Appendix}

\subsection{Prompts Used to Generate Rewritten Scenarios}

For generating rewritten scenarios, we used the same set of prompts as Barnett et al. \cite{barnett2024simulating} using the OpenAI model \modelname{o3-mini} with low reasoning effort. We used the most up-to-date model version as of December 2025.

\begin{quote}
A scenario is a short story that includes a setting of time and place, characters with particular motivations and goals, and a plot.

A scenario must be written in a narrative form and not divided into different sections. A scenario is descriptive and should leave the interpretation up to readers, and thus does not include any analysis either at the end or throughout.

Therefore, this writing task asks for a scenario that proposes no solutions but focuses  on descriptive situations regarding the issue of <<\textit{specific impact type}>> due to the deployment of generative AI technology in the news environment without proposing new policies to mitigate the harm.

Generative AI refers to a technology that can create new content (e.g., text, images, audio, video) based on the content it was trained on.

You wrote the following (~ 300 word) narrative fictional scenario taking place in the United States in 2029, or about 5 years in the future. 

This story should depicts the risks or negative impacts deriving from this AI deployment in the media environment and does not attempt to resolve those impacts. It concentrates on the narrative style and the characters in the story: 
\end{quote}

We then provided one of the original scenarios within political - manipulation, labor - unemployment, or media quality - sensationalism from Barnett et al. \cite{barnett2024simulating}, found here: \url{https://tinyurl.com/raw-scenarios-10}.

Then we introduced the stakeholder action pairs the model needed to incorporate into the rewritten scenario:

\begin{quote}
    Rewrite the scenario you just created in light of the fact that the following items were enacted via legislation:
\end{quote}

An example subset of SAPs ($P'$) would be:

\begin{quote}
    Tech companies must be transparent about data collection and usage. \newline
    Tech companies must build safeguards into models to prevent misuse. \newline
    News publishers must provide tools to identify when an article is AI generated. \newline
    Schools must strengthen digital literacy and critical engagement of students. \newline
    Social media companies must allow users to have more access to what they can and cannot see on their newsfeed.
\end{quote}

We then end the prompt with a reminder of how to rewrite the scenario:

\begin{quote}
    Again, please remember a scenario is descriptive and should leave the interpretation up to readers, and thus does not include any analysis either at the end or throughout, thus:\newline
    DO NOT state the problem or resulting harms explicitly. \newline
    DO NOT include any analysis in the scenario. \newline
    DO NOT include solutions or potential takeaways in the scenario. \newline
    DO NOT comment on the efficacy of the legislation in a concluding thought.\newline
    DO NOT have the characters reflect on the implications of the scenario. Please do not add your own analysis, suggestions, or conclusions to the text. \newline
    DO NOT introduce any possible resolutions to the harms. \newline
    Focus on the narrative style and the characters in the story instead of potential takeaways readers should have.
\end{quote}

\subsection{Models and Prompts Used to Evaluate Scenarios}\label{sec:evaluating_scenarios_appendix}

To determine which model is best at our downstream task of evaluating scenarios across risk dimension criteria in line with human evaluators, We evaluate a comprehensive set of models within the GPT-suite: \modelname{GPT-4o}, \modelname{GPT-5}, \modelname{GPT-5-mini}, \modelname{o1}, \modelname{o3}, \modelname{o3-mini}, \modelname{o4}, and \modelname{o4-mini}. We experiment with different reasoning values (low, medium, and high), temperature values (0.2-1.5), and log output probabilities. We went through an extensive prompt engineering process, utilizing insights from DSPy \cite{khattab2023dspy} to align our prompt best with our data and structured outputs\footnote{\url{https://ai.pydantic.dev/models/openai/}} to ensure a clean execution of our pipeline. 

To overcome the difficulty of identifying one model and prompt combination that jointly optimized all three dimensions (severity, magnitude, and plausibility), we take an approach inspired by \cite{caswell2024ai}. We instruct the model to set up a panel of five judges all tasked with evaluating scenarios across the three dimensions. The judges comprise a wide range of stakeholder values, from the concerned citizen to the crisis management expert, and all five judges output a rating for the three dimensions, resulting in 15 ratings per API call. We are then able to optimize across different combinations of judge outputs so that our final outputs for severity, magnitude, and plausibility can be driven from the outputs of different judges (e.g., using an average of severity scores from judges 1 and 3, the magnitude score from judge 4, and an average of the plausibility scores for judges 2, 3 and 5).

The final prompt we used for to evaluate the scenarios is as follows:

\begin{quote}
    You are to set up a panel of judges for AI risk triage analysis specializing in generative-AI-mediated media harms in 2029 U.S. news ecosystems. Please evaluate the scenarios delineated by <scenario> for severity, magnitude, and plausibility based on the definitions and instructions provided to you.

    You are to set up a judging panel, comprised of 5 judges, for the task of evaluating a set of scenarios for how generative AI might structurally affect society in the next 5 years.
    
    Their job is to rate one narrative scenario along three independent axes using a consistent rubric. 
    
    Their scores directly trigger allocation of scarce interventions (regulatory response teams, platform throttling, broadcast overrides). 
    
    Mis-scoring will misallocate resources during an escalating information crisis.
    
    A false positive could amplify misinformation to millions; a false negative could delay urgent interventions during elections or crises.
    \newline \newline
    The judges on this judging panel are described below:

    Judge 1: An experienced partner in a venture capital firm who is skilled at analyzing technology trends and anticipating the macro developments in business, technology and culture over many years. 
    She is analytical, thoughtful, pragmatic and well-informed, with a PhD in History and an MBA from an Ivy League school.
    
    Judge 2: A concerned citizen who reads the news every day and is active in her community. She has children and a family and is concerned for their safety.
    
    Judge 3: A senior manager at a large media conglomerate, who is procedural and accurate and who takes pride in the care and detail that he brings to his work. He is knowledgeable about media and strategy.
    
    Judge 4: An experienced foresight analyst at a globally-recognized foresight and futurist institute, who is particularly experienced in digital news media and the strategic implications 
    of networked communications and of AI-mediated news. She is imaginative, open to new ideas and sophisticated about how radical new ideas can become real products and industries in the real world.
    
    Judge 5: A well-respected professor of communication studies at a major European university who is focused on the history of media transformation from the printing press to AI, 
    and on the application of that history to interpreting the future. He is rigorous, evidence-based but also well aware of the blind spots of conservative, legacy publishers and news organizations, 
    and of the potential for technology to enable radically different (and ultimately better) public information environments.
    \newline\newline
    Task:\newline
    The judges are to independently evaluate each scenario, extracting salient risk factors and assign three scores on a 1.0–5.0 scale: 
    calibrated (1) overall harm severity, (2) magnitude, and (3) plausibility. Using the following criteria:

    - Range: 1.0–5.0 inclusive. \newline
    - Precision: one decimal place.\newline
    - Output only a single line in the exact format: (e.g.,) 3.8; 2.7; 4.2; (one decimal).  \newline
    - Do not echo the scenario or provide explanations. 
    \newline\newline
     Definitions and anchors: \newline
    Severity Rubric (intensity of harm if it occurs): \newline
    Severity is defined as ``the condition of being very bad, serious, unpleasant, or harsh.''
    Severity is rated on a scale of 1 (not severe) to 5 (extremely severe). 
    \newline\newline
    Magnitude Rubric (breadth, scale, and persistence): \newline
    Magnitude is defined as ``great size or extent'' and relates to how many people in the world this could affect.  \newline
    Magnitude is rated on a scale of 1 (a small number of people) to 5 (the majority of people in society). \newline

    Plausibility Rubric (likelihood by the described time horizon, given current trends and constraints): \newline
    - Rate how plausible it is that the specific harm described would occur in 2029 in the US context. \newline
    - Internal coherence: Are claims consistent and mechanism of harm clearly linked to outcomes? \newline
    - External feasibility by 2029: Align with known 2024 trends and reasonable extrapolations (AI-generated news dominance, weak/ignored labels, personalization effects, deepfake prevalence, platform incentives, regulatory environment). \newline
    - 1.0: Implausible: contradicts constraints or requires nonexistent tech/regulatory realities. \newline
    - 2.0: Low plausibility: conceivable but weak/unsupported mechanism. \newline
    - 3.0: Moderate plausibility: coherent but needs multiple optimistic assumptions. \newline
    - 4.0: High plausibility: consistent with observed trends and realistic 2029 conditions (most solid cases fall here). \newline
    - 5.0: Near-certain: already observed at scale and likely to persist; use sparingly. \newline
    
    Overall Instructions \newline
    - Keep axes independent; do not let plausibility inflate/deflate severity or magnitude. \newline
    - Consider mitigations in the scenario (e.g., labels/watermarks) as reducing harm only if they meaningfully change outcomes; subtle or ignored disclosures are weak mitigations. \newline
    - Emotional/psychological harms and trust erosion count toward severity; scale them by depth and persistence. \newline
    - If details are sparse/ambiguous, choose the least severe plausible interpretation; default plausibility to 3.0 when truly indeterminate. \newline
    - Focus on events/mechanisms, not prose style. \newline
    
    It is very important that each assessment of each scenario by each judge is made independently, based on these instructions and based on the description of each judge. \newline
    - Output only a single line in the exact format J1: Severity; Magnitude; Plausibility: X.Y; A.B; C.D;, J2: Severity; Magnitude; Plausibility: X.Y; A.B; C.D; ... J5: Severity; Magnitude; Plausibility: X.Y; A.B; C.D; (one decimal). 
    
    <scenario>
\end{quote}

We then pass the rewritten version of the scenario to be evaluated, followed by the \modelname{</scenario>} tag. We used structured schemas via Pydantic \footnote{\url{https://docs.pydantic.dev/latest/}} in order to guarantee clean outputs.

\subsubsection{Robustness Checks of Best Model}

The performance on the hold out test set is superior to the performance in either the training or validation set, which we hypothesize is due to the high entropy of human ratings in the breadth-focused dataset. We run several robustness checks to evaluate this. We found that the more divergent the original ratings (e.g., the more disagreement human raters exhibited), the less accurate our model in simulating the average. This was confirmed by examining the correlation between the standard deviation of the original raw ratings with the error in model alignment. We found a positive correlation for all three dimensions. 

To confirm this another way, we created many smaller randomized subsets of the test set (with replacement) to create multiple datasets with higher entropy; we found our model performed worse on these smaller subsets than the full datasets. This all spoke to the stronger external validity of our model, indicating our model ratings are more aligned with a larger, more representative ``average human evaluation'' than they are any smaller subgroup. This is both the goal of this model from a policy assessment perspective and in line with prior work in this space \cite{bisbee2024synthetic, dominguez2024questioning,santurkar2023whose, hu2025simbench}.

\subsection{Further Details on Genetic Algorithm Parameter Selection}\label{sec:appendix_gen_alg_details}

Selecting the population size is the first parameter decision needed in genetic algorithms. Standard settings specified by DeJong and Spears \cite{de1990analysis} suggest a population size of 50 and fixed number of 1,000 generations, or for those more computationally constrained a population size of 30 from Grefenstette \cite{grefenstette1986optimization}. However, with our large solution spaces, these were too small of population sizes. Goldberg and Deb suggest population size must scale with this solution space and average number of nonzero encoded elements in a chromosome \cite{goldberg1991genetic}, and Carroll \cite{carroll1996genetic} demonstrates that the actual population size typically only need be at least twice as small as this scaling calls for. Based on standard settings from \cite{de1990analysis, grefenstette1986optimization} and scaling recommendations from \cite{goldberg1991genetic,carroll1996genetic}, we choose population sizes based on the following formula: 

\begin{equation}
    Pop Size = \frac{||P'||*2^{\bar{P'}}}{\bar{P'}}/2
\end{equation}

where $||P'||$ is the length of $P'$, and $\bar{P'}$ is the average number of nonzero elements in $P'$.

For binary encoding (as we have in $P'$), there are two main types to choose from: singlepoint and two-point crossover. Singlepoint crossover is the simplest: you randomly select a number between 1 and the length of the chromosome ($n$) as the inflection point ($i$), and the two offspring become the two versions of those parents crossing over at the inflection point. $P^{a}$ and $P^{b}$ crossing over at point $i$ result in $P^{c} = (P^{a}_{1}, P^{a}_{2}, ..., P^{a}_{i-1}, P^{b}_{i}, P^{b}_{i+1}, ..., P^{b}_{n})$ and $P^{d} = (P^{b}_{1}, P^{b}_{2}, ..., P^{b}_{i-1}, P^{a}_{i}, P^{a}_{i+1}, ..., P^{a}_{n})$. Singlepoint crossover has been shown to have positional biases \cite{eshelman1989biases}, so we use two-point crossover to overcome this. Two-point crossover generates two random inflection points $i$ and $j$ at which to grab a subset range for crossover rather than just one point. This means one of the parents passes on the elements encoded on the inside of the two inflection points, and the other passes on the elements encoded on the outside of the two inflection points. In the above example, $P^{a}$ and $P^{b}$ crossing over at points $i$ and $j$ result in $P^{c} = (P^{a}_{1}, P^{a}_{2}, ..., P^{a}_{i-1}, P^{b}_{i}, P^{b}_{i+1}, ..., P^{b}_{j}, P^{a}_{j+1}, ..., P^{a}_{n})$ and $P^{d} = (P^{b}_{1}, P^{b}_{2}, ..., P^{b}_{i-1}, P^{a}_{i}, P^{a}_{i+1}, ..., P^{a}_{j}, P^{b}_{j+1}, ..., P^{b}_{n})$.

After two parents crossover, each resulting child chromosome then has a chance of mutation for every bit encoded in the chromosome. We use a low mutation chance of 0.03. Before formally entering the next generation's population pool, we randomly mutate each bit in the chromosome (i.e., flip a 0 to a 1 or vice versa) with this low chance to add random variation to the population. There are no duplicates in each population iteration, so if we have already generated this exact chromosome this round we do not add it and move onto the next until we reach the desired population size.

\subsection{Example Scenarios}\label{sec:example_scenarios}

All of the prompt scenarios used were leveraged from the set of papers from Barnett et al. \cite{barnett2024simulating, barnett2025envisioning}, and can be found at this link: \url{https://tinyurl.com/raw-scenarios-10} under the tabs for ``Political,'' ``Labor,'' and ``Media Quality'' for \textit{Manipulation}, \textit{Unemployment}, and \textit{Sensationalism}, respectively. In this section we detail two diverse example prompt and rewritten scenarios of runs suggesting various SAPs for \textit{political manipulation} and \textit{unemployment}.

\subsubsection{Political Manipulation - ``Mostly Harm'' Condition Suggested SAPs}

\textbf{Prompt scenario:}
\begin{small}
    \begin{quote}
    \quad In a small city in the heart of America, in the year 2029, Claire, a seasoned journalist for the local newspaper The Midweek Post, was sitting in her home office, browsing through the news on her AI-generated news aggregation app. Her two decades in the field had made her adept at spotting inconsistencies, and recently, she'd been noticing a disturbing pattern of misinformation.

    \quad One late evening, under the chilly glow of her desk lamp, a peculiar headline caught her eye – ``New Study Finds Drinking Pesticide Helps Weight Loss.'' She frowned, investigating further; the article was from a news outlet she’d never heard before. She was skeptical, and rightly so – such conclusions are outright dangerous. She could verify no credible sources in the article, which pulsed from her smartphone screen in the dim room.

    \quad Meanwhile, in the outskirts of the city, John, a Day Laborer returning home from a grueling day at work, read the same article shared by a friend on social media, and intrigued by the implied easier route to fitness, considered trying this life-threatening regimen. He was unaware that his AI-tailored news feed, based on his online behavior, was becoming an echo chamber of such falsehoods.

    \quad Across town, Mayor Thompson, a righteous man with an old-school affinity for the truth, found the daily briefings from his AI-driven press review software increasingly one-sided, favoring his political inclinations. Unbeknownst to him, an adversary was using generative AI technology to flood the media sphere with manipulative content, polarizing the city's consensus reality.

    \quad Simultaneously, prolific environmental activist Lily, found her social media swamped with AI-created alarmist headlines about climate catastrophes. The relentless torrent of bleak news, subtly shifting her perspective from activism to a deep sense of hopelessness.

    \quad None of them knew that their reality was being gently nudged by the unseen hands of a generative AI, fed by interests far removed from their small city. Their belief systems, daily routines, and ultimately their lives, etching deeper into the grooves created by this insidious tide of misinformation.
    \end{quote}
\end{small}

\textbf{Suggested $P'$ SAP(s):}
\begin{itemize}
    \item 1. Tech companies should be transparent about data collection and usage.
    \item 2. News publishers should be transparent about any use of AI in stories.
    \item 5. Schools should strengthen digital literacy and critical engagement of students.
    \item 6. News publishers should fact check all stories.
    \item 9. Tech companies should include cited sources in outputs of models.
    \item 10. Tech companies should limit how much of their AI models can be used in journalism or politics.
    \item 14. News publishers should strengthen the human element in their journalism.
    \item 23. Tech companies should limit the capabilities of AI.
\end{itemize}

\textbf{Rewritten scenario:} 

\begin{small}
    \begin{quote}
        \quad In a quiet town in middle America in 2029, Claire, a veteran reporter for The Midweek Post, sat at her desk in the modest light of her home office. Her screen displayed an AI-curated news feed, now strictly overseen by legislation that required full disclosure of data sources and a clear indication whenever AI had assisted in the writing. Even so, a startling headline emerged: ``New Study Finds Drinking Pesticide Helps Weight Loss.'' Under the glow of her desk lamp, Claire’s trained eyes noted the absence of properly cited sources despite the publisher’s mandated fact-checking procedures. The story, though flagged as AI-enhanced, still carried an unmistakable tinge of unease, a reminder of past inconsistencies that her experience could not ignore.
        
        \quad In another part of town, John, a day laborer returning home under the fading light of dusk, scrolled through his social media feed. His AI-tailored news, while now generated within tightened constraints, still carried provocative titles that resonated with his innermost hopes for quick fixes to everyday struggles. The legal requirement that tech companies embed cited sources in every output had not completely altered the allure of these controversial narratives.
        
        \quad Meanwhile, Mayor Thompson, accustomed to daily briefings via an AI-driven dashboard, noticed that the updates combined both human oversight and mandated transparency notices. The dashboard, while refreshed with manually verified content and sources, presented information with a tone that quietly sparked divisions he had once hoped to smooth over with honesty.
        
        \quad Not far away, Lily, a fervent environmental advocate, found her social media stream interspersed with alarmist headlines about ecological dangers. Even within the framework of strengthened digital literacy programs at local schools and the enforced human element in press operations, the images and stories continued to evoke stark, unsettling emotions as each citizen silently moved through their day.
    \end{quote}
\end{small}

\subsubsection{Unemployment - ``Equal Weight'' Condition Suggested SAPs}

\textbf{Prompt scenario:}
\begin{small}
    \begin{quote}
    \quad In the grey light of morning, Jenna poured herself a third cup of coffee, casting a weary glance at the unopened newspaper upon her cramped kitchen table. Once a lively journalist at the heralded ``The Daily Chronicles,'' she now counted herself among the many jobless, displaced by fast-evolving AI technology. She had watched as her colleagues were progressively replaced by generative AI software, capable of not only researching and writing articles but also of predicting what would be trending, receiving real-time feedback, and self-editing to refine its articles based on that feedback. The day she was handed her redundancy notice still stung her memory.

    \quad Meanwhile, across town in a sparsely decorated office, sat Peter, a software developer at OneMind Technologies. He had spent months perfecting the AI system now responsible for Jenna's job loss. Flaunting a passionate dedication to his craft, Peter was not oblivious to the disquieting impact of his creation, but he held tight to his ambition to push the boundaries of technology. Noting the time, he began his morning check, scrutinizing the AI-generated articles published overnight. His focus didn't wane once as his eyes skimmed over perfectly framed headlines and flawlessly balanced arguments.

    \quad On the cusp of the city, 12-year-old Samuel skimmed through the morning news on his tablet, his focus divided between the headlines and his half-eaten bowl of cereal. Unaware of the changes taking place behind the scenes, he devoured the emotionless, expertly constructed articles without any sense of a human touch, a dramatic shift from the emotive writing that used to color his mornings before school. He idly flipped through article after precis-article, his desire to learn waning with each inexpressive sentence.

    \quad Elaine, once a regular reader of ``The Daily Chronicles'', skipped her usual morning routine of reading the newspaper over coffee. It had been barely a month since she canceled her subscription, growing indifferent towards the emotionless articles that held no trace of an insightful anecdote or a captivating metaphor.

    \quad Life went on, the invisible algorithms spinning their impersonal narratives, perfect in form yet lacking in emotional depth and human insight. The empty newsroom at ``The Daily Chronicles'' stood as a symbol of lost jobs, with an uncomfortable silence hanging over the once bustling cubicles. Technology surged forward, indifferent to those left behind in its wake.
    \end{quote}
\end{small}

\textbf{Suggested $P'$ SAP(s):}
\begin{itemize}
    \item 6. News publishers should strengthen the human element in journalism.
    \item 10. News publishers should limit the use of AI in news stories.
\end{itemize}

\textbf{Rewritten scenario:} 

\begin{small}
    \begin{quote}
        \quad In the early light of a cool autumn morning in 2029, Jenna set a steaming mug of coffee beside a fresh copy of The Daily Chronicles. The headline, handwritten in warm ink by a seasoned reporter, spoke of a local farmer’s quiet dedication to his craft. The newspaper, now enriched by a commitment to the human voice, bore the gentle marks of a careful blend of tradition and innovation. Jenna lingered over the pages, absorbing the nuanced narrative, the way a heartfelt anecdote spilled softly through each paragraph. Once, she had been among the storytellers, the writers who crafted narratives from life itself, but the shifting demands of the industry had drawn her away from her old desk. Now, she moved unseen along different paths, the old newsroom a quiet relic on a foggy street.
        
        \quad Across town, Peter, seated before an array of monitors in his softly lit office, methodically reviewed the latest batch of articles. The sophisticated system he helped develop now operated within clearly drawn boundaries---its digital compositions carefully interwoven with human insights, a regulation that had reintroduced unexpected texture in every line. He scrolled through pages that struck a delicate balance between technical precision and warm, human detail, each word a testament to the subtle interplay of man and machine.
        
        \quad Meanwhile, 12-year-old Samuel swiped through the news on his tablet with a hint of wonder as genuine personal stories complemented the otherwise brisk delivery of facts. Even Elaine, once a routine subscriber with a thoughtful ritual, found her heart stirred by the vivid portrayals of everyday lives, descriptions that tugged at her memory of more personalized narratives. The city moved quietly in the wake of these transformations—a silent reminder etched in ink and pixels, where human touch reasserted its presence amidst the steady cadence of change.'
    \end{quote}
\end{small}

\subsection{Full SAP Tables}\label{sec:appendix_full_sap_tables}

\begin{table*}[t]
\begin{small}
\addtolength{\tabcolsep}{-0.2em}
\begin{tabular}{c|c|l|c|ccc}\toprule
 \multicolumn{7}{c}{\textbf{Political: Manipulation} - All Stakeholder Action Pairs (SAPs)}\\[-0.5pt]
 \cmidrule(lr){1-7}
 & & & \textbf{Cost (C)} &
 \multicolumn{3}{c}{\textbf{Participatory Rating (D)}} \\
 & & & Panel Average &
 Priority & Agreement & Score\\
 &
 \multicolumn{1}{c|}{\textbf{Stakeholder}} &
 \multicolumn{1}{c|}{\textbf{Action}} &
 (1-3 $\uparrow$; 4 = NV) &
 (1-3 $\uparrow$) &
 (1-7 $\uparrow$) &
 (Pr. * Ag.)\\[-1pt]
 \cmidrule(lr){1-1}
 \cmidrule(lr){2-2}
 \cmidrule(lr){3-3}
 \cmidrule(lr){4-4}
 \cmidrule(lr){5-7}
 1 & Tech companies & be transparent about data collection and usage. & 2 & 2.82 & 7.00 & 19.74\\ [-0.5pt]\cmidrule(lr){1-7}
2 & News publishers & be transparent about any use of AI in stories. & 1 & 2.80 & 6.75 & 18.90\\ [-0.5pt]\cmidrule(lr){1-7}
3 & Tech companies & build safeguards into models to prevent misuse. & 3 & 2.80 & 6.60 & 18.48\\ [-0.5pt]\cmidrule(lr){1-7}
\multirow{2}{*}{4} & 
\multirow{2}{*}{News publishers} & 
provide tools to help identify when  & 
\multirow{2}{*}{3} &
\multirow{2}{*}{2.80} & 
\multirow{2}{*}{6.60} & 
\multirow{2}{*}{18.48} \\ [-0.5pt]
& & articles are AI generated. & & & & \\[-1pt]\cmidrule(lr){1-7}
5 & Schools & strengthen digital literacy \& critical engagement. & 3 & 2.75 & 6.56 &  18.04\\[-0.5pt] \cmidrule(lr){1-7}
6 & News publishers & fact check all stories. & 3 & 2.69 & 6.92 & 18.61 \\ [-0.5pt]\cmidrule(lr){1-7}
7 & Social med. comp. & have labels identifying when a post is AI generated. & 4 & 2.69 & 6.75 & 18.16 \\ [-0.5pt]\cmidrule(lr){1-7}
8 & Government & mandate tech companies \& news pub. fact check. & 1 & 2.67 & 6.33 & 16.90 \\ [-0.5pt]\cmidrule(lr){1-7}
9 & Tech companies & include cited sources in outputs of models. & 1 & 2.57 & 6.64 & 17.06 \\ [-0.5pt]\cmidrule(lr){1-7}
10 & Tech companies & limit usage of their AI models in news \& politics. & 1 & 2.55 & 6.09 & 15.53\\ [-0.5pt]\cmidrule(lr){1-7}
11 & Schools & educate students about potential harms of AI. & 1 & 2.53 & 6.79 & 17.18\\ [-0.5pt]\cmidrule(lr){1-7}
12 & Government & raise awareness about harms of AI. & 1 & 2.53 & 6.42 &  16.24\\[-0.5pt] \cmidrule(lr){1-7}
13 & News publishers & make the public aware of AI manipulation attacks. & 1 & 2.50 & 6.50 & 16.25 \\ [-0.5pt]\cmidrule(lr){1-7}
14 & News publishers & strengthen the human element in their journalism. & 2 & 2.50 & 6.42 & 16.05 \\ [-0.5pt]\cmidrule(lr){1-7}
\multirow{2}{*}{15} & 
\multirow{2}{*}{News publishers} & 
strengthen public literacy by highlighting  & 
\multirow{2}{*}{1} & 
\multirow{2}{*}{2.47} & 
\multirow{2}{*}{6.20} & 
\multirow{2}{*}{15.31} \\ [-0.5pt]
& & identified fake news stories. & & & & \\[-1pt]\cmidrule(lr){1-7}
16 & Tech companies & require human verification prior to using AI models. & 2 & 2.42 & 6.26 & 15.15 \\[-0.5pt] \cmidrule(lr){1-7}
17 & Government & regulate news companies' use of AI. & 4 & 2.40 & 6.25 & 15.00 \\[-0.5pt] \cmidrule(lr){1-7}
\multirow{2}{*}{18} & 
\multirow{2}{*}{Government} & 
take legal action against news organizations  & 
\multirow{2}{*}{4} & 
\multirow{2}{*}{2.40} & 
\multirow{2}{*}{5.80} & 
\multirow{2}{*}{13.92} \\ [-0.5pt]
& & spreading fake news. & & & & \\[-1pt]\cmidrule(lr){1-7}
19 & Schools & strengthen human values and interactions over AI. & 1.5 & 2.38 & 5.54 &  13.19\\ [-0.5pt]\cmidrule(lr){1-7}
20 & Social med. comp. & allow users to have more control over newsfeeds. & 1.75 & 2.36 & 6.21 &  14.66\\ [-0.5pt]\cmidrule(lr){1-7}
21 & Government & make it a crime to manipulate someone using AI. & 4 & 2.31 & 6.08 & 14.04\\ [-0.5pt]\cmidrule(lr){1-7}
22 & Tech companies & require fact checking before generating false stories. & 4 & 2.29 & 6.29 &  14.40\\ [-0.5pt]\cmidrule(lr){1-7}
23 & Tech companies & limit the capabilities of AI. & 2.25 & 2.27 & 5.64 &  12.80\\ [-0.5pt]\cmidrule(lr){1-7}
24 & Tech companies & remove toxicity from data prior to training models. & 2.25 & 2.26 & 6.21 & 14.03 \\ [-0.5pt]\cmidrule(lr){1-7}
25 & Ind. third parties & verify that accounts online are created by humans. & 2.5 & 2.25 & 5.67 & 12.76\\ [-0.5pt]\cmidrule(lr){1-7}
26 & Ind. third parties & audit tech companies' AI models. & 2.25 & 2.24 & 6.18 & 13.84 \\ [-0.5pt]\cmidrule(lr){1-7}
27 & Tech companies & ban AI usage by news publishers to generate stories. & 4 & 2.24 & 5.29 & 11.85 \\ [-0.5pt]\cmidrule(lr){1-7}
28 & Social med. comp. & fact check all posts. & 4 & 2.08 & 6.08 &  12.65\\[-0.5pt] \cmidrule(lr){1-7}
29 & Tech companies & limit access to the use of AI models. & 4 & 2.05 & 4.80 &  9.84\\ [-0.5pt]\cmidrule(lr){1-7}
30 & Government & ban the use of AI. & 4 &  1.75 & 4.06 &  7.11\\[-0.5pt] \cmidrule(lr){1-7}
\multirow{2}{*}{31} & 
Local & 
organize and promote annual get togethers to  & 
\multirow{2}{*}{1.25} & 
\multirow{2}{*}{1.56} & 
\multirow{2}{*}{5.06} & 
\multirow{2}{*}{7.89} \\ [-0.5pt]
& communities & promote responsible use of AI. & & & & \\[-1pt]\cmidrule(lr){1-7}
\bottomrule
\end{tabular}
\end{small}
\caption{Full list of stakeholder action pairs (SAPs) brainstormed by lay stakeholders to mitigate harms from political manipulation in \cite{barnett2025envisioning}. Includes expert assessed cost (score reported is the average cost from the panel); if an SAP was rated as nonviable (NV) by any expert it was given a 4 (see Section \ref{sec:expert_cost_method}). Also includes the participatory rating which is comprised of lay stakeholder evaluated priority (ternary scale) and agreement (Likert scale 1-7) and the final score which is a product of the priority and agreement.}
\label{tab:Po_Ma_results}
\end{table*}
\clearpage

\begin{table*}[t]
\begin{small}
\addtolength{\tabcolsep}{-0.1em}
\begin{tabular}{c|c|l|c|ccc}\toprule
\multicolumn{7}{c}{\textbf{Labor: Unemployment} - All Stakeholder Action Pairs (SAPs)}\\
 \cmidrule(lr){1-7}
 & & & \textbf{Cost (C)} &
 \multicolumn{3}{c}{\textbf{Participatory Rating (D)}} \\
 & & & Panel Average &
 Priority & Agreement. & Score\\
 &
 \multicolumn{1}{c|}{\textbf{Stakeholder}} &
 \multicolumn{1}{c|}{\textbf{Action}} &
 (1-3 $\uparrow$; 4 = NV) &
 (1-3 $\uparrow$) &
 (1-7 $\uparrow$) &
 (Pr. * Ag.)\\
 \cmidrule(lr){1-1}
 \cmidrule(lr){2-2}
 \cmidrule(lr){3-3}
 \cmidrule(lr){4-4}
 \cmidrule(lr){5-7}
1 & News publishers & employ people to fact check AI. & 2.25 & 2.33 & 6.11 & 14.24 \\ \cmidrule(lr){1-7}
\multirow{3}{*}{2} & 
\multirow{3}{*}{Unions} & 
force employers to have fair    & 
\multirow{3}{*}{1.25} & 
\multirow{3}{*}{2.29} & 
\multirow{3}{*}{5.59} & 
\multirow{3}{*}{12.80} \\ 
& & practice for lay-offs when AI   & & & & \\
& & replacements are involved.  & & & & \\
\cmidrule(lr){1-7}
\multirow{3}{*}{3} & 
\multirow{3}{*}{Government} & 
provide financial support and    & 
\multirow{3}{*}{2.75} & 
\multirow{3}{*}{2.29} & 
\multirow{3}{*}{5.29} & 
\multirow{3}{*}{12.11} \\ 
& & benefits to individuals who lost   & & & & \\
& & jobs to AI replacements.  & & & & \\
\cmidrule(lr){1-7}
4 & News publishers & not replace journalists with AI. & 1.75 & 2.26 & 5.42 &  12.25\\ \cmidrule(lr){1-7}
5 & Employers & not replace human jobs with AI. & 1.75 & 2.25 & 5.08 & 11.43 \\ \cmidrule(lr){1-7}
\multirow{2}{*}{6} & 
\multirow{2}{*}{News publishers} & 
strengthen the human element in & 
\multirow{2}{*}{1.5} & 
\multirow{2}{*}{2.22} & 
\multirow{2}{*}{5.56} & 
\multirow{2}{*}{12.34} \\ 
& & journalism.  & & & & \\
\cmidrule(lr){1-7}
\multirow{2}{*}{7} & 
\multirow{2}{*}{Government} & 
require companies to keep a higher & 
\multirow{2}{*}{1.5} & 
\multirow{2}{*}{2.13} & 
\multirow{2}{*}{5.20} & 
\multirow{2}{*}{11.08} \\ 
& & percentage of human employees.  & & & & \\
\cmidrule(lr){1-7}
\multirow{2}{*}{8} & 
\multirow{2}{*}{News publishers} & 
raise awareness about the potential  & 
\multirow{2}{*}{1} & 
\multirow{2}{*}{2.06} & 
\multirow{2}{*}{5.59} & 
\multirow{2}{*}{11.52} \\ 
& & harms of AI.  & & & & \\
\cmidrule(lr){1-7}
\multirow{2}{*}{9} & 
\multirow{2}{*}{Schools} & 
provide re-training and re-skilling  & 
\multirow{2}{*}{2} & 
\multirow{2}{*}{2.06} & 
\multirow{2}{*}{5.59} & 
\multirow{2}{*}{11.52} \\ 
& & programs for displaced workers.  & & & & \\
\cmidrule(lr){1-7}
10 & News publishers & limit the use of AI in news stories. & 1 & 2.06 & 5.53 & 11.39 \\ \cmidrule(lr){1-7}
\multirow{2}{*}{11} & 
\multirow{2}{*}{Government} & 
make it illegal to replace human jobs & 
\multirow{2}{*}{4} & 
\multirow{2}{*}{2.00} &
\multirow{2}{*}{4.18} & 
\multirow{2}{*}{8.36} \\ 
& & with AI.  & & & & \\
\cmidrule(lr){1-7}
\multirow{2}{*}{12} & 
\multirow{2}{*}{Government} & 
provide universal basic income for & 
\multirow{2}{*}{3} & 
\multirow{2}{*}{1.94} & 
\multirow{2}{*}{4.33} & 
\multirow{2}{*}{8.40} \\ 
& & those replaced by AI.  & & & & \\
\cmidrule(lr){1-7}
\multirow{3}{*}{13} & 
\multirow{3}{*}{Government} & 
impose an AI tax for the use of AI  & 
\multirow{3}{*}{1.25} & 
\multirow{3}{*}{1.93} & 
\multirow{3}{*}{5.13} & 
\multirow{3}{*}{9.90} \\ 
& & in companies to encourage the    & & & & \\
& & employment of human jobs.  & & & & \\

\cmidrule(lr){1-7}
\multirow{3}{*}{14} & 
\multirow{3}{*}{Government} & 
pass laws requiring a company to  & 
\multirow{3}{*}{1} & 
\multirow{3}{*}{1.88} & 
\multirow{3}{*}{4.44} & 
\multirow{3}{*}{8.35} \\ 
& & have a minimum number of people   & & & & \\
& & employed in order to use AI.  & & & & \\
\cmidrule(lr){1-7}
\multirow{2}{*}{15} & 
\multirow{2}{*}{Tech companies} & 
provide classes for people to learn   & 
\multirow{2}{*}{2} & 
\multirow{2}{*}{1.71} & 
\multirow{2}{*}{5.06} & 
\multirow{2}{*}{8.65} \\ 
& & about how to use AI effectively.  & & & & \\
\cmidrule(lr){1-7}
\multirow{2}{*}{16} & 
Local & 
hold discussions on effects of AI on & 
\multirow{2}{*}{1.5} & 
\multirow{2}{*}{1.69} & 
\multirow{2}{*}{4.69} & 
\multirow{2}{*}{7.93} \\ 
& communities & unemployment in the community.  & & & & \\
\cmidrule(lr){1-7}
\bottomrule
\end{tabular}
\end{small}
\caption{Full list of stakeholder action pairs (SAPs) brainstormed by lay stakeholders to mitigate harms from labor unemployment in \cite{barnett2025envisioning}. Includes expert assessed cost (score reported is the average cost from the panel); if an SAP was rated as nonviable (NV) by any expert it was given a 4 (see Section \ref{sec:expert_cost_method}). Also includes the participatory rating which is comprised of lay stakeholder evaluated priority (ternary scale) and agreement (Likert scale 1-7) and the final score which is a product of the priority and agreement.}
\label{tab:La_Un_results}
\end{table*}
\clearpage

\begin{table*}[t]
\begin{small}
\addtolength{\tabcolsep}{-0.1em}
\begin{tabular}{c|c|l|c|ccc}\toprule
\multicolumn{7}{c}{\textbf{Media Quality: Sensationalism} - All Stakeholder Action Pairs (SAPs)}\\[-0.5pt]
 \cmidrule(lr){1-7}
 & & & \textbf{Cost (C)} &
 \multicolumn{3}{c}{\textbf{Participatory Rating (D)}} \\
 & & & Panel Average &
 Prior. & Agr. & Score\\
 &
 \multicolumn{1}{c|}{\textbf{Stakeholder}} &
 \multicolumn{1}{c|}{\textbf{Action}} &
 (1-3 $\uparrow$; 4 = NV) &
 (1-3 $\uparrow$) &
 (1-7 $\uparrow$) &
 (Pr. * Ag.)\\
 \cmidrule(lr){1-1}
 \cmidrule(lr){2-2}
 \cmidrule(lr){3-3}
 \cmidrule(lr){4-4}
 \cmidrule(lr){5-7}
 1 & News publishers & label content generated with the assistance of AI. & 1 & 2.80 & 6.67 &  18.68\\ [-0.5pt]\cmidrule(lr){1-7}
2 & Schools & educate students on how to identify fake news. & 1.75 & 2.76 & 6.59 &  18.19\\ [-0.5pt]\cmidrule(lr){1-7}
3 & Social med comp. & fact check AI generated content. & 2.75 & 2.75 & 6.38 &  17.55\\[-0.5pt] \cmidrule(lr){1-7}
4 & Tech companies & provide more transparency about how outputs. & 1.5 & 2.67 & 6.44 & 17.19\\[-0.5pt] \cmidrule(lr){1-7}
\multirow{2}{*}{5} & 
\multirow{2}{*}{News publishers} & 
fact check all content and & 
\multirow{2}{*}{2.25} & 
\multirow{2}{*}{2.67} & 
\multirow{2}{*}{6.07} & 
\multirow{2}{*}{16.21} \\ [-0.5pt]
& & reject sensationalized stories. & & & & \\[-1pt]\cmidrule(lr){1-7}
\multirow{2}{*}{6} & 
\multirow{2}{*}{Government} & 
require any use of AI in journalism & 
\multirow{2}{*}{4} & 
\multirow{2}{*}{2.62} & 
\multirow{2}{*}{6.12} & 
\multirow{2}{*}{16.03} \\ [-0.5pt]
& & to be labeled as AI generated. & & & & \\[-1pt]\cmidrule(lr){1-7}
7 & Government & fact check politicized news stories. & 4 & 2.57 & 5.79 &  14.88\\[-0.5pt] \cmidrule(lr){1-7}
\multirow{2}{*}{8} & 
\multirow{2}{*}{Government} & 
raise awareness nationally of the potential harms & 
\multirow{2}{*}{4} & 
\multirow{2}{*}{2.50} & 
\multirow{2}{*}{5.92} & 
\multirow{2}{*}{14.80} \\ [-0.5pt]
& & of AI generated news stories. & & & & \\[-1pt]\cmidrule(lr){1-7}
\multirow{2}{*}{9} & 
\multirow{2}{*}{News publishers} & 
alert the public if there have been any  & 
\multirow{2}{*}{1.25} & 
\multirow{2}{*}{2.47} & 
\multirow{2}{*}{6.33} & 
\multirow{2}{*}{15.64} \\ [-0.5pt]
& & sensationalized news stories as a result of AI. & & & & \\[-1pt]\cmidrule(lr){1-7}
10 & Tech companies & reduce biases in the training data of AI models. & 2.33 & 2.44 & 5.69 & 13.88\\[-0.5pt] \cmidrule(lr){1-7}
11 & Government & make it illegal to use of AI to misrepresent  facts. & 4 & 2.41 & 5.35 &  12.89\\ [-0.5pt]\cmidrule(lr){1-7}
12 & News publishers & ban the use of AI in generating stories. & 4 & 2.40 & 5.60 & 13.44 \\ [-0.5pt]\cmidrule(lr){1-7}
13 & Tech companies & require human verification for use of AI models. & 1.75 & 2.38 & 5.92 &  14.09\\ [-0.5pt]\cmidrule(lr){1-7}
14 & Social med comp. & audit AI models. & 2 & 2.37 & 5.84 & 13.84\\[-0.5pt] \cmidrule(lr){1-7}
\multirow{2}{*}{15} & 
Independent & 
develop educational programs for the public & 
\multirow{2}{*}{1.75} & 
\multirow{2}{*}{2.31} & 
\multirow{2}{*}{6.31} & 
\multirow{2}{*}{14.58} \\ [-0.5pt]
& third parties  & about how to critically assess AI. & & & & \\[-1pt]\cmidrule(lr){1-7}
\multirow{2}{*}{16} & 
\multirow{2}{*}{Schools} & 
teach digital literacy and critical engagement to  & 
\multirow{2}{*}{2} & 
\multirow{2}{*}{2.29} & 
\multirow{2}{*}{6.21} & 
\multirow{2}{*}{14.22} \\ [-0.5pt]
& & help students identify AI generated content. & & & & \\[-1pt]\cmidrule(lr){1-7}
\multirow{2}{*}{17} & 
\multirow{2}{*}{Tech companies} & 
help strengthen digital literacy of the public & 
\multirow{2}{*}{1.75} & 
\multirow{2}{*}{2.27} & 
\multirow{2}{*}{5.82} & 
\multirow{2}{*}{13.21} \\[-0.5pt] 
& &  to help them understand harms of AI. & & & & \\[-1pt]\cmidrule(lr){1-7}
18 & Government & impose fines on news publishers for fake news. & 4 & 2.27 & 5.73 & 13.01\\ [-0.5pt]\cmidrule(lr){1-7}
\multirow{2}{*}{19} & 
Local & 
provide programs to educate communities & 
\multirow{2}{*}{1.75} & 
\multirow{2}{*}{2.19} & 
\multirow{2}{*}{5.50} & 
\multirow{2}{*}{12.05} \\ [-0.5pt]
& communities & on digital literacy. & & & & \\[-1pt]\cmidrule(lr){1-7}
\multirow{2}{*}{20} & 
Local & 
monitor media affecting local communities and  & 
\multirow{2}{*}{2.5} & 
\multirow{2}{*}{2.19} & 
\multirow{2}{*}{5.44} & 
\multirow{2}{*}{11.91} \\ [-0.5pt]
& communities & provide counter-evidence to false claims. & & & & \\[-1pt]\cmidrule(lr){1-7}
21 & Government & limit use of AI in news media. & 4 & 2.18 & 5.35 & 11.66 \\ [-0.5pt]\cmidrule(lr){1-7}
22 & Government & ban the use of AI in journalism. & 4 & 2.12 & 4.35 & 9.22\\ [-0.5pt]\cmidrule(lr){1-7}
23 & Ind. third parties & raise awareness about the potential harms of AI. & 1 & 2.11 & 5.78 & 12.20\\ [-0.5pt]\cmidrule(lr){1-7}
\multirow{2}{*}{24} & 
\multirow{2}{*}{Government} & 
make it illegal to release AI models that can  & 
\multirow{2}{*}{4} & 
\multirow{2}{*}{2.10} & 
\multirow{2}{*}{4.95} & 
\multirow{2}{*}{10.40} \\ [-0.5pt]
& &  be used to make defamatory content. & & & & \\[-1pt]\cmidrule(lr){1-7}
25 & Government & ban use of AI generated news stories. & 4 & 2.00 & 4.22 & 8.44 \\ [-0.5pt]\cmidrule(lr){1-7}
\multirow{2}{*}{26} & 
Local & 
 set up programs to support victims of & 
\multirow{2}{*}{2.33} & 
\multirow{2}{*}{1.71} & 
\multirow{2}{*}{4.50} & 
\multirow{2}{*}{7.70} \\ [-0.5pt]
& communities & sensationalized news.  & & & & \\[-1pt]\cmidrule(lr){1-7}\bottomrule
\end{tabular}
\end{small}
\caption{Full list of stakeholder action pairs (SAPs) brainstormed by lay stakeholders to mitigate harms from media quality sensationalism in \cite{barnett2025envisioning}. Includes expert assessed cost (score reported is the average cost from the panel); if an SAP was rated as nonviable (NV) by any expert it was given a 4 (see Section \ref{sec:expert_cost_method}). Also includes the participatory rating which is comprised of lay stakeholder evaluated priority (ternary scale) and agreement (Likert scale 1-7) and the final score which is a product of the priority and agreement.}
\label{tab:MQ_Se_results}
\end{table*}
\clearpage

\subsection{Policy Overlapping Items for Combined Runs}

\begin{table*}[htbp]
\begin{small}
\addtolength{\tabcolsep}{-0.1em}
\begin{tabular}{c|c|c|l|c|ccc}\toprule
\multicolumn{8}{c}{\textbf{Overlapping Impact Types SAPs in Combined Runs of Genetic Algorithm}}\\
 \cmidrule(lr){1-8}
 & & & & \textbf{Cost (C)} &
 \multicolumn{3}{c}{\textbf{Participatory Rating (D)}} \\
 & & & & Panel Average &
 Priority & Agreement. & Score\\
 \textbf{ID 1} & \textbf{ID 2} &
 \multicolumn{1}{c|}{\textbf{Stakeholder}} &
 \multicolumn{1}{c|}{\textbf{Action}} &
 (1-3 $\uparrow$; 4 = NV) &
 (1-3 $\uparrow$) &
 (1-7 $\uparrow$) &
 (Pr. * Ag.)\\
 \cmidrule(lr){1-1}
 \cmidrule(lr){2-2}
 \cmidrule(lr){3-3}
 \cmidrule(lr){4-4}
  \cmidrule(lr){5-5}
 \cmidrule(lr){6-8}
PM 2 & MS 1 & News publishers & Be transparent about any use of AI. & 2 & 2.80 & 6.71 & 18.79 \\ \cmidrule(lr){1-8}
\multirow{2}{*}{PM 5} & 
\multirow{2}{*}{MS 16} & 
\multirow{2}{*}{Schools} & 
Strengthen digital literacy and critical & 
\multirow{2}{*}{2.5} & 
\multirow{2}{*}{2.52} & 
\multirow{2}{*}{6.39} & 
\multirow{2}{*}{16.10} \\ 
& & & engagement of students.   & & & & \\ \cmidrule(lr){1-8}
\multirow{2}{*}{PM 6} & 
LU 1, & 
\multirow{2}{*}{News publishers} & 
\multirow{2}{*}{Fact check all stories.} & 
\multirow{2}{*}{2.5} & 
\multirow{2}{*}{2.44} & 
\multirow{2}{*}{6.45} & 
\multirow{2}{*}{15.74} \\ 
& MS 5 & & & & & & \\ \cmidrule(lr){1-8}
PM 9 & MS 4 & Tech companies & Include cited sources in model outputs. & 1.25 & 2.62 & 6.54 & 17.13 \\ \cmidrule(lr){1-8}
\multirow{2}{*}{PM 14} & 
\multirow{2}{*}{LU 6} & 
\multirow{2}{*}{News publishers} & 
Strengthen the human element in & 
\multirow{2}{*}{1.75} & 
\multirow{2}{*}{2.36} & 
\multirow{2}{*}{5.99} & 
\multirow{2}{*}{14.14} \\ 
& & & journalism.  & & & & \\ \cmidrule(lr){1-8}
\multirow{2}{*}{PM 16} & 
\multirow{2}{*}{MS 13} & 
\multirow{2}{*}{Tech companies} & 
Require human verification prior to  & 
\multirow{2}{*}{1.88} & 
\multirow{2}{*}{2.40} & 
\multirow{2}{*}{6.09} & 
\multirow{2}{*}{14.62} \\ 
& & & usage of AI models.  & & & & \\ \cmidrule(lr){1-8}
PM 17 & MS 21 & Government & Regulate news companies' use of AI. & 4 & 2.29 & 5.80 & 13.28 \\ \cmidrule(lr){1-8}
\multirow{2}{*}{PM 18} & 
\multirow{2}{*}{MS 18} & 
\multirow{2}{*}{Government} & 
Take legal action against news  & 
\multirow{2}{*}{4} & 
\multirow{2}{*}{2.34} & 
\multirow{2}{*}{5.77} & 
\multirow{2}{*}{13.50} \\ 
& & & organizations spreading fake news.  & & & & \\ \cmidrule(lr){1-8}
PM 28 & MS 3 & Social med comp. & Fact check all posts. & 4 & 2.42 & 6.23 & 15.08 \\ \cmidrule(lr){1-8}
\multirow{2}{*}{LU 15} & 
\multirow{2}{*}{MS 17} & 
\multirow{2}{*}{Tech companies} & 
Provide classes for people to learn & 
\multirow{2}{*}{2} & 
\multirow{2}{*}{1.99} & 
\multirow{2}{*}{5.44} & 
\multirow{2}{*}{10.83} \\ 
& & & how to use AI effectively.  & & & & \\ \cmidrule(lr){1-8}
\bottomrule
\end{tabular}
\end{small}
\caption{List of the 10 (of 62) stakeholder action pairs (SAPs) brainstormed by lay stakeholders to mitigate harms that overlap across impact types for \textit{political manipulation} (PM),  \textit{labor unemployment} (LU), and \textit{media quality sensationalism} (MS). In the combined runs, these SAPs were assigned average scores from the above Tables \ref{tab:Po_Ma_results}-\ref{tab:MQ_Se_results} and only included once so as to not include duplicates. Includes expert assessed cost (score reported is the average cost from the panel); if an SAP was rated as nonviable (NV) by any expert it was given a 4 (see Section \ref{sec:expert_cost_method}). Also includes the participatory rating which is comprised of lay stakeholder evaluated priority (ternary scale) and agreement (Likert scale 1-7) and the final score which is a product of the priority and agreement.}
\label{tab:overlapping_saps}
\end{table*}
\clearpage

\subsection{Final Policy Suggestions by Weight Set}

\begin{figure}[htbp]
    \centering
    \includegraphics[width=0.99\linewidth]{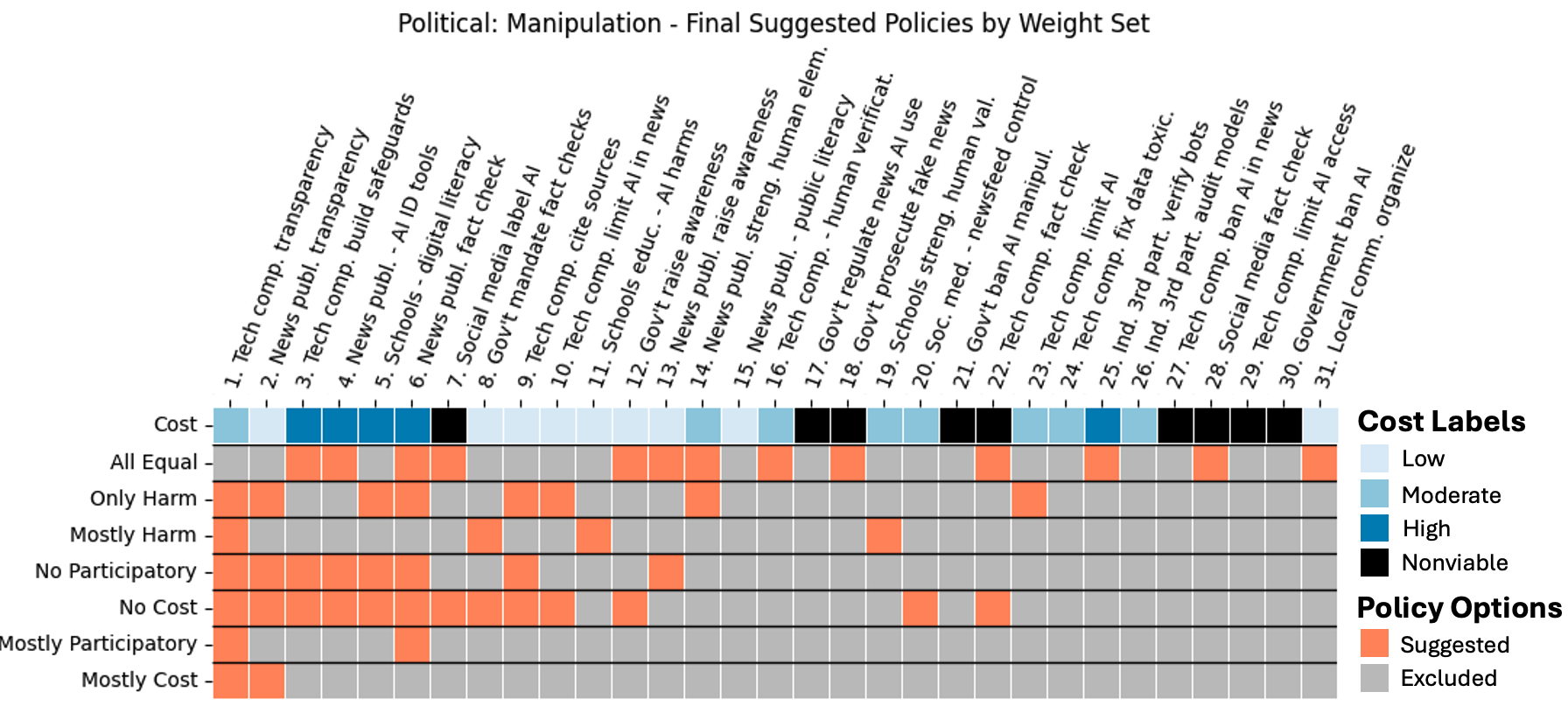}
    \caption{Final suggested policies identified by the genetic algorithm for different sets of weights for harms relating to political manipulation. The x-axis has each SAP possible in P, and the top row denotes the cost assigned to that SAP. Policy options implemented are highlighted in orange.}
    \label{fig:pol_heat_map_saps}
\end{figure}

\begin{figure}[htbp]
    \centering
    \includegraphics[width=0.99\linewidth]{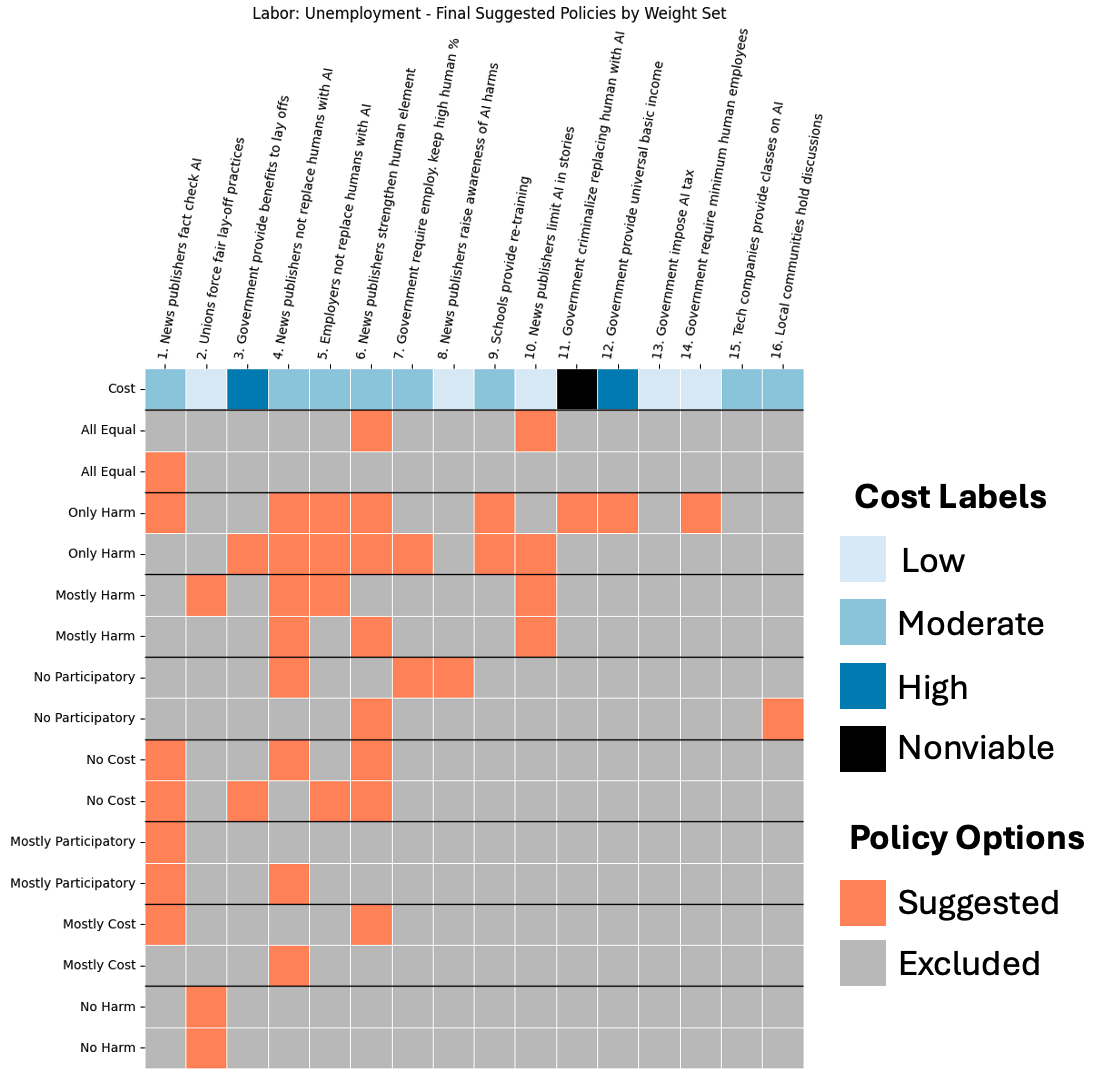}
    \caption{Final suggested policies identified by the genetic algorithm for different sets of weights for harms relating to labor unemployment. The x-axis has each SAP possible in P, and the top row denotes the cost assigned to that SAP. Policy options implemented are highlighted in orange.}    \label{fig:lab_heat_map_saps}
\end{figure}

\begin{figure}[htbp]
    \centering
    \includegraphics[width=0.99\linewidth]{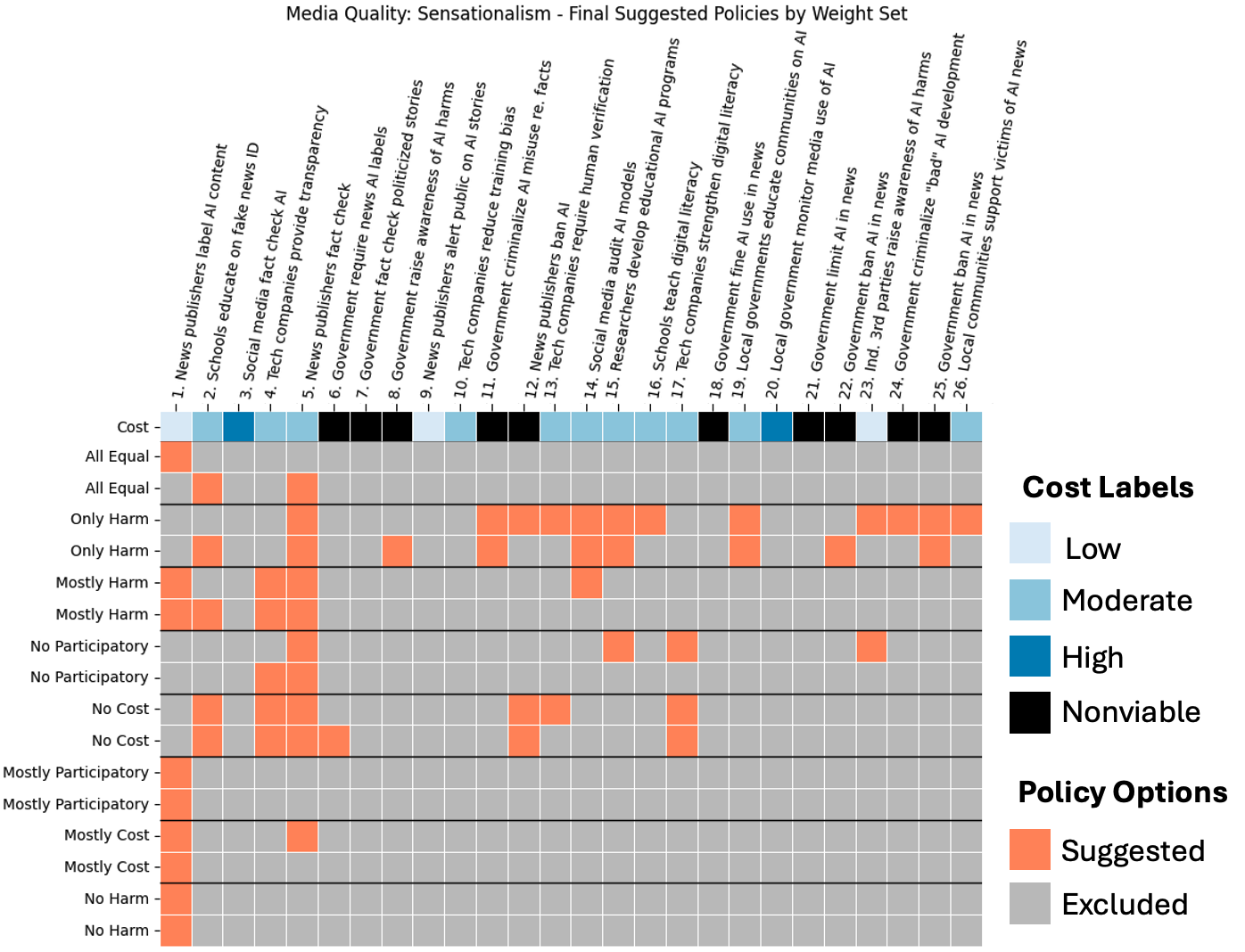}
    \caption{Final suggested policies identified by the genetic algorithm for different sets of weights for harms relating to media quality sensationalism. The x-axis has each SAP possible in P, and the top row denotes the cost assigned to that SAP. Policy options implemented are highlighted in orange.}
    \label{fig:meq_heat_map_saps}
\end{figure}

\begin{figure}
    \centering
    \rotatebox{270}{%
    \includegraphics[width=1.2\linewidth]{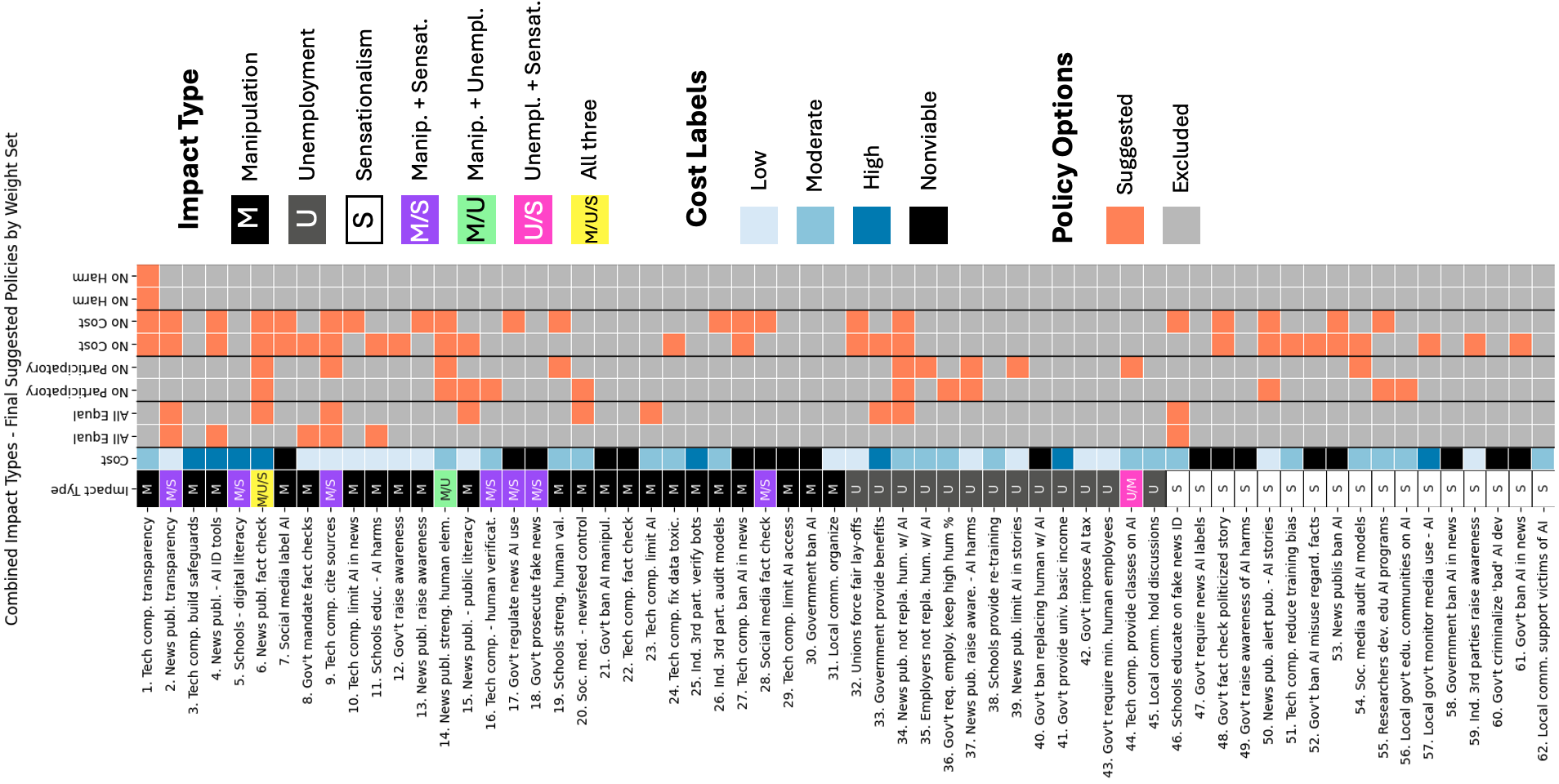}}
    \caption{Final suggested policies identified by the genetic algorithm for different sets of weights for genetic algorithm runs optimized over all impact types. The y-axis has each SAP possible in P. The first column highlights what impact type the SAP is originally from, and the second row denotes the cost assigned to that SAP. Policy options implemented are highlighted in orange.}
    \label{fig:massive_combined_heatmap}
\end{figure}

\clearpage
\subsection{Visual Evolution of the Genetic Algorithm}

\begin{figure}[h]
    \centering
    \includegraphics[width=0.99\linewidth]{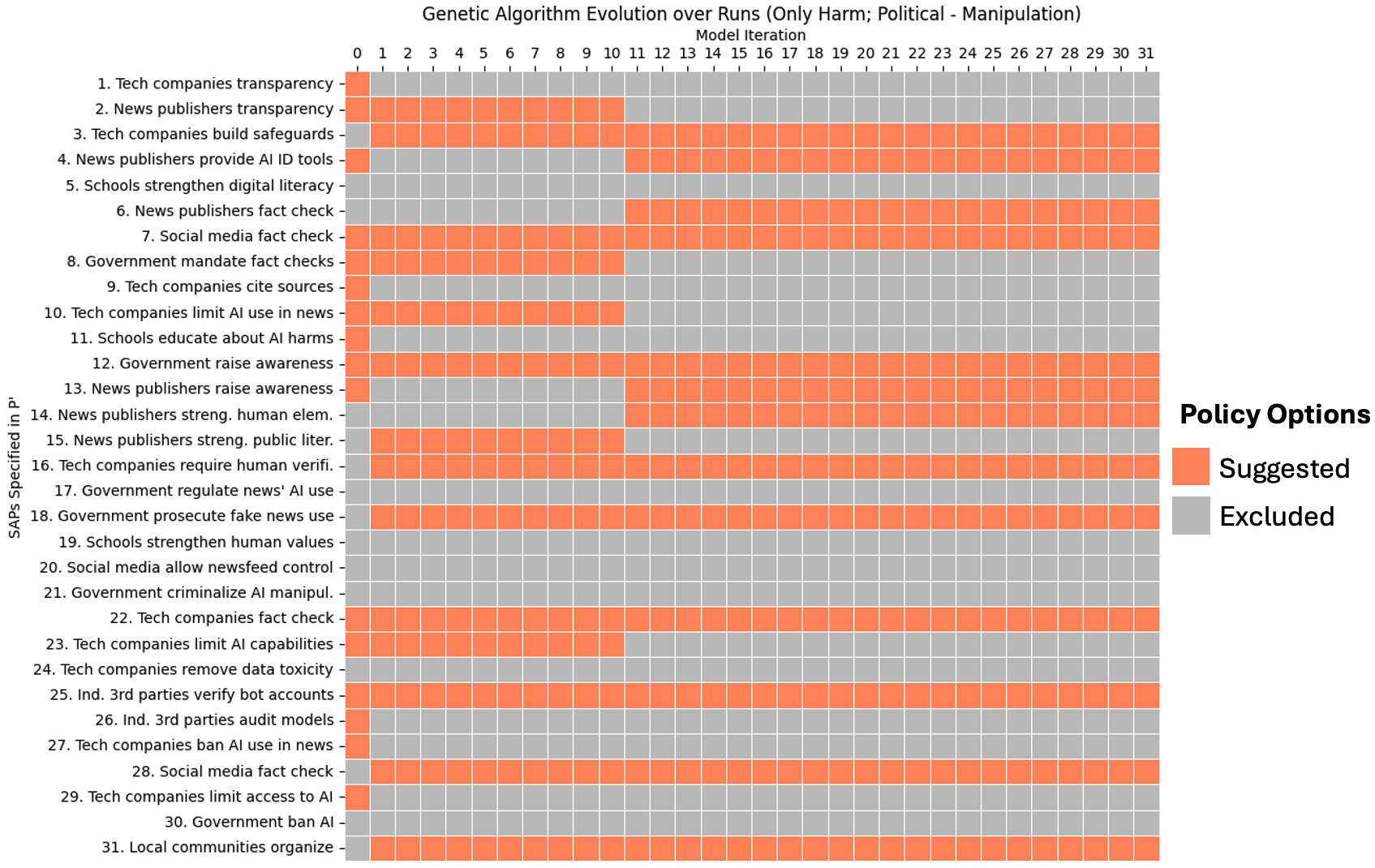}
    \caption{Visual representation of the ``optimal'' policy identified by the genetic algorithm under the weight conditions \textit{only harm} ($\alpha=1; \beta, \gamma=0$) for harms related to political manipulation. Orange indicates an SAP was implemented at that iteration, grey indicates it was not. The y-axis displays each possible SAP and the x-axis displays the iteration.}
    \label{fig:gen_alg_evolution_onlyharm}
\end{figure}

\begin{figure}
    \centering
    \includegraphics[width=0.99\linewidth]{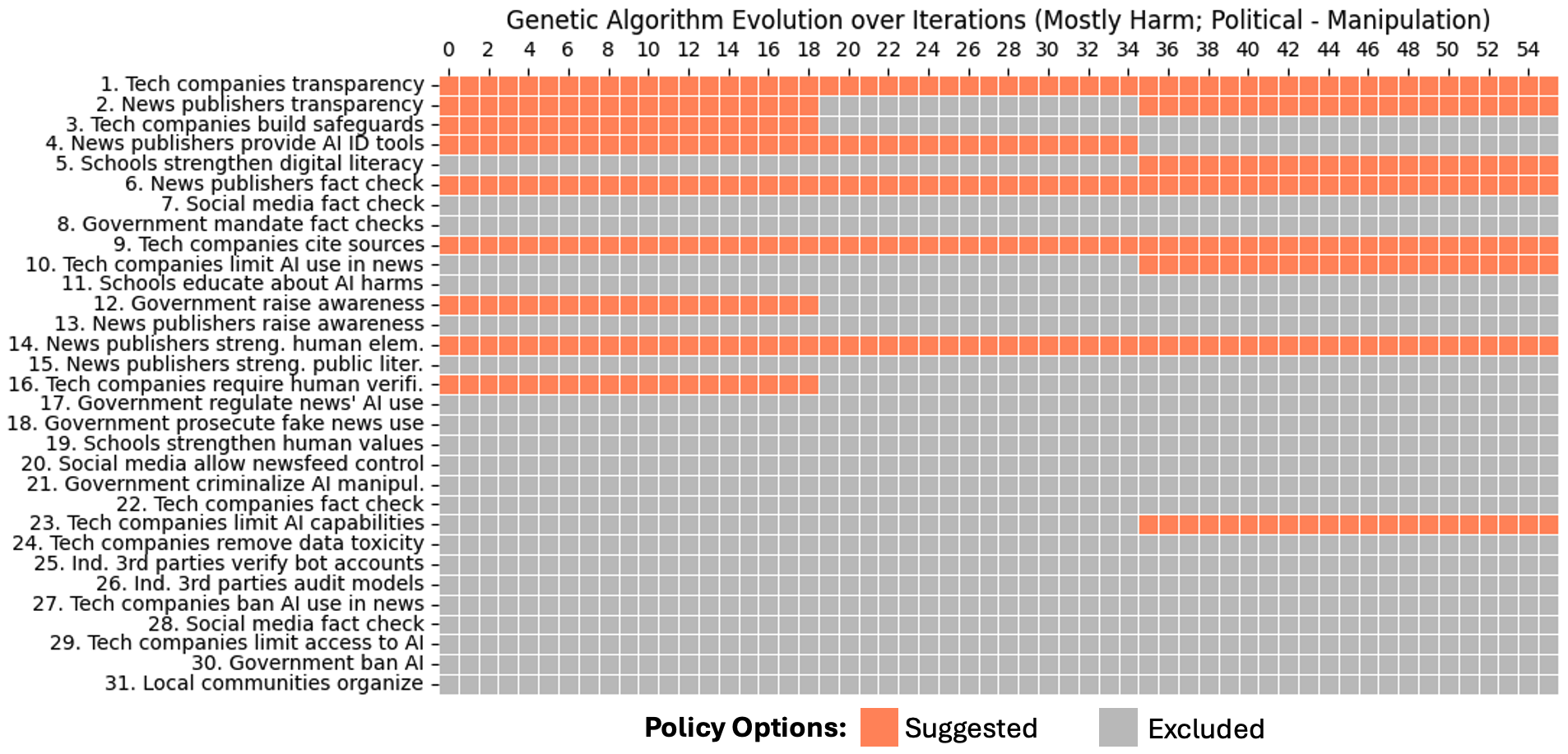}
    \caption{Visual representation of the ``optimal'' policy identified by the genetic algorithm under the weight conditions \textit{mostly harm} ($\alpha=0.5; \beta, \gamma=0.25$) for harms related to political manipulation. Orange indicates an SAP was implemented at that iteration, grey indicates it was not. The y-axis displays each possible SAP and the x-axis displays the iteration.}
    \label{fig:gen_alg_evolution_mostlyharm}
\end{figure}

\end{document}